\documentclass[aps,prd,amsmath,amssymb,twocolumn,superscriptaddress,preprintnumbers,nofootinbib]{revtex4-1}
\pdfoutput=1
\usepackage{graphicx}
\usepackage{amsthm}
\usepackage{amsmath}
\usepackage{graphicx,subfigure}% Include figure files
\usepackage{dcolumn}% Align table columns on decimal point
\usepackage{bm}% bold math
\usepackage{slashed}

\newcommand{\be}{\begin{eqnarray}}
\newcommand{\ee}{\end{eqnarray}}
\newcommand{\bea}{\begin{eqnarray}}
\newcommand{\eea}{\end{eqnarray}}

\newcommand{\keV}{{~\rm keV}}

\newcommand{\GeV}{{~\rm GeV}}

\newcommand{\mX}{m_{_\chi} }

\newcommand{\mN}{m_{_N}}

\newcommand{\muX}{\mu_{_\chi}}
\newcommand{\muN}{\mu_{_N}}

\newcommand{\muXg}{\mu_{_{\chi\gamma}}}

\newcommand{\ER}{E_{_R}}

\newcommand{\sDD}{\sigma_{_{\rm  DD}}}
\newcommand{\sDZ}{\sigma_{_{\rm  DZ}}}
\newcommand{\ZZ}{{\rm Z}}

\newcommand{\SUWeak}{{\rm SU_{_W}(2)}}

\newcommand{\medf}{\psi}
\newcommand{\meds}{\varphi}
\newcommand{\mmedf}{M_f}
\newcommand{\mmeds}{M_s}
\newcommand{\MSdiv}{\frac{1}{\epsilon}}
\newcommand{\mOverM}{\zeta_M}
\begin{document}

\title{Magnetic dipole moment of neutral particles from quantum corrections at two-loop order}
\author{Carlos Tamarit}
\email{ctamarit@perimeterinstitute.ca}
\affiliation{Perimeter Institute for Theoretical Physics 31 Caroline St. N, Waterloo, Ontario, Canada N2L 2Y5.}
\author{Itay Yavin}
\email{iyavin@perimeterinstitute.ca}
\affiliation{Department of Physics \& Astronomy, McMaster University 1280 Main St. W. Hamilton, Ontario, Canada, L8S 4L8.}
\affiliation{Perimeter Institute for Theoretical Physics 31 Caroline St. N, Waterloo, Ontario, Canada N2L 2Y5.}

%\date{\today}

\begin{abstract}
The tentative gamma-ray line in the Fermi data at $\sim 135$ GeV motivates a dark matter candidate that couples to photons through loops of charged messengers. It was recently shown that this model can explain the observed line, but achieving the correct phenomenology requires a fairly sizable coupling between the WIMP and the charged messengers. While strong coupling by itself is not a problem, it is natural to wonder whether the phenomenological success is not spoiled by higher order quantum corrections. In this work we compute the dominant two-loop contributions to the electromagnetic form-factors of the WIMP and show that over a large portion of the relevant parameter space these corrections are under control and the phenomenology is not adversely affected. We also discuss more generally the effects of these form-factors on signals in direct-detection experiments as well as on the production of the WIMP candidate in colliders.  In particular, for low masses of the charged messengers the production rate at the LHC enjoys an enhancement from the threshold singularity associated with these charged states.
\end{abstract}

\maketitle

%%%%%%%%%%%%%%%%
% Introduction
%%%%%%%%%%%%%%%%
\section{Introduction}
\label{sec:intro}

As is well-known, neutral particles uncharged under the electromagnetic field may still participate in one-photon exchange processes via a dipole transition. For a relativistic neutral fermion, $\chi$, the interaction is written as,
\be
\label{eqn:MiDMInteraction}
\mathcal{L}_{\rm dipole}= \left(\frac{\muX}{2}\right)\bar \chi^* \sigma_{\mu\nu} F^{\mu\nu} \chi + \mathrm{h.c.},
\ee
where throughout we use natural units with $\alpha = e^2/4\pi$, and here $\mu_\chi$ is the magnetic dipole strength, $F^{\mu\nu}$ is the electromagnetic field-strength tensor, and $\sigma_{\mu\nu} = i[\gamma_\mu,\gamma_\nu]/2$ is the commutator of two Dirac matrices. This interaction vanishes when $\chi = \chi^*$ is a single Majorana fermion, but otherwise describes a dipole transition between two distinct fermions $\chi$ and $\chi^*$ as in the case of neutrinos~\cite{Kusenko:2009up}, or between the two Majorana components of a single Dirac fermion $\chi = \chi^*$. If whatever constitutes the observed dark matter in the Universe is a new fundamental particle, then its eponymous character implies charged neutrality. Nevertheless, dark matter may still have a magnetic dipole moment as in (\ref{eqn:MiDMInteraction}) and the consequences of this interaction have been explored in a variety of contexts~\cite{Bagnasco:1993st,Pospelov:2000bq,Sigurdson:2004zp,Gardner:2008yn, Masso:2009mu,Cho:2010br,An:2010kc,McDermott:2010pa,Chang:2010en,Banks:2010eh,Goodman:2010qn,Fortin:2011hv,DelNobile:2012tx}. In this case, the dark matter $\chi$ is a type of a weakly interacting massive particle (WIMP) and we refer to this scenario as magnetic dark matter (MDM). 

The recent observation of an excess of gamma-ray events from the center of the galaxy in the Fermi satellite data strengthened the motivation for considering the magnetic dipole interactions of dark matter; see refs.~\cite{Ackermann:2012qk, Bringmann:2012vr,Weniger:2012tx,Tempel:2012ey,Su:2012ft} for early investigations of the line, refs.~\cite{Hektor:2012jc,Hektor:2012kc,Hektor:2012ev,Whiteson:2012hr,Finkbeiner:2012ez,Rao:2012fh} for more detailed follow-ups, and refs.~\cite{Buchmuller:2012rc,Cohen:2012me,Cholis:2012fb,Blanchet:2012vq,Asano:2012zv} for model-independent constraints on continuum emissions. The connection between the magnetic dipole interaction and the Fermi line was first made explicit in ref.~\cite{Weiner:2012cb}, where in addition the importance of the Rayleigh interaction involving a two-photon coupling was emphasized and clarified (see also the later work of ref.~\cite{Tulin:2012uq}).  

At length scales much smaller than the dipole strength $\muX$, the interaction~(\ref{eqn:MiDMInteraction}) is no longer an appropriate description of the physics and some ultraviolet (UV) completion of the theory is needed. If the neutral particle $\chi$ is in fact a composite made of smaller constituents which are themselves charged, then the dipole interaction arises simply from the charge separation and the dipole strength $\muX$ is related to the compositeness length-scale times the charge of the constituents. The neutron is the prime archetype with a dipole moment of $-1.9 \muN$, where $\muN$ is the nuclear magneton. A model of MDM along these lines was constructed in ref~\cite{Cline:2012bz} and later investigated in ref.~\cite{Dissauer:2012xa} as well. 

\begin{figure}
\begin{center}
\includegraphics[width=0.23 \textwidth]{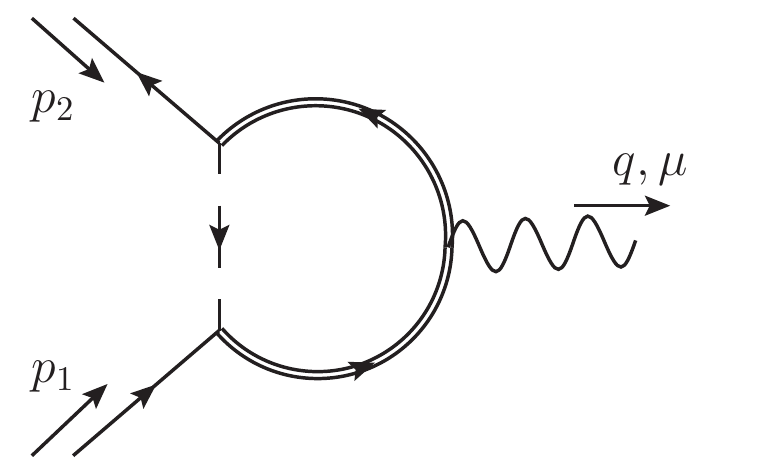}
\includegraphics[width=0.23 \textwidth]{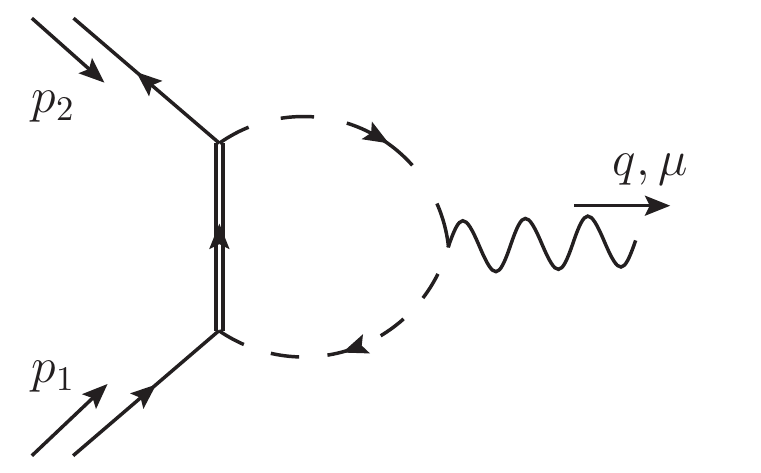}
\end{center}
\caption{Magnetic dipole operator generated at one-loop.}
\label{fig:dipole_loop}
\end{figure}

However, the WIMP $\chi$ does not have to be a composite particle as the interaction (\ref{eqn:MiDMInteraction}) may arise from quantum corrections even for elementary particles. This idea was first made explicit in the context of MDM in ref.~\cite{Weiner:2012cb}. The theory does require new charged states, and in ref.~\cite{Weiner:2012gm} the authors considered a simple UV completion including a new charged scalar ($\meds$) and a charged fermion ($\medf$) in ${\bf 2}_\frac{1}{2}$ representations of the electroweak group $SU(2)_W\times U(1)_Y$ which couple to the WIMP through a Yukawa interaction
\be
\label{eqn:Yukawa-type}
\mathcal{L}\supset \lambda\, \bar{\medf}\chi \meds.
\ee
To lowest order in perturbation theory, this Yukawa interaction generates a magnetic dipole moment for the WIMP through the diagrams shown in Fig.~\ref{fig:dipole_loop}. In ref.~\cite{Weiner:2012gm} it was shown that this leads to successful phenomenology in the context of the Fermi-line, but the Yukawa coupling required is large $\lambda \sim \sqrt{4\pi}$. It is important to recognize that strong coupling does not immediately imply the breakdown of perturbation theory. Nevertheless, one is legitimately concerned that the next order corrections are large and may spoil the phenomenological success of the theory. Thus, in this paper, we set to compute the next order (two-loop) corrections to the interaction~(\ref{eqn:MiDMInteraction}) due to the new Yukawa coupling. We will show that the next order corrections are indeed under control and do not adversely modify the phenomenology associated with thermal freeze-out. In addition, we work out the effects of the electromagnetic form-factors on the signal in direct-detection and collider experiments.

The paper is organized as follows: In the next section we present the relevant diagrams and the different ingredients involved in the computation; in section~\ref{sec:results} we present the results; section~\ref{sec:pheno} is devoted to the resulting phenomenology; we conclude in section~\ref{sec:conclusions}. The reader interested only in the effects on the phenomenology of this model may wish to skip directly to section~\ref{sec:pheno} or even the conclusions. 

\section{Setup}
\label{sec:Setup}

We consider the most general renormalizable Lagrangian for the WIMP $\chi$, the charged scalar $\meds$ and the charged fermion $\medf$. The fields $\meds$ and $\medf$ are chosen to be SU(2) doublets with hypercharge $1/2$, and we will refer to them as ``messengers''. The Lagrangian is
\be
\nonumber
\mathcal{L} &=& \bar{\chi}\left( i\slashed{\partial} - \mX\right)\chi -\frac{1}{2}\delta \, \chi C \chi + \bar{\medf}\left( i\slashed{D} - \mmedf\right)\medf 
\\
\nonumber
&+& \left(D^\mu\meds \right)^\dag D_\mu\meds - \mmeds^2 \meds^\dag \meds
\\
&+& \lambda \bar{\medf}\chi \meds - \frac{\kappa}{4} \left| \meds^\dag \meds \right|^2 + {\rm h.c.} ,
\label{eqn:Lagrangian}
\ee
where $D_\mu = \partial_\mu - i g W_\mu^a\tau^a - i Y g' B_\mu$ is the covariant derivative associated with the $\SUWeak \times U_Y(1)$ gauge-bosons, $W_\mu^a$ and $B_\mu$, respectively, and $\tau^a$ are the $\SUWeak$ generators obeying tr$\left(\tau^a\tau^b\right) = \tfrac{1}{2} \delta^{ab}$ and related to the Pauli matrices through $\tau^a=\tfrac{1}{2} \sigma^a$. The parameter $\delta$ is a Majorana mass which splits the Dirac state into a pseudo-Dirac pair when $\delta \ll \mX$. In this work we concentrate exclusively on this possibility. The quartic coupling $\kappa$ is inconsequential to the phenomenology at one-loop, but it does contribute at the two-loop level. When SM fields are included, other operators appear at the renormalizable level. Those were discussed extensively in ref.~\cite{Liu:2013gba} and their effect on the phenomenology was addressed in detail. For the purpose of the present paper we note that these operators do not present an "in principle" obstruction for successful phenomenology, although their inclusion certainly needs to be done with care. More importantly, unless their size is so large as to contribute significantly at tree-level (an undesirable feature), their effects at loop-level will be entirely subdominant compared to the contributions we considered herein and so we neglect their effect in the rest of this paper.

An effective coupling between the WIMP and hypercharge of the form $\bar \chi \Gamma^\mu \chi B_\mu$ is generated through loops involving $\medf$ and $\meds$ as in Fig.~\ref{fig:dipole_loop}. The most general one-gauge-boson interaction vertex consistent with Lorentz symmetry depends only on the momentum transfer $q^2$, and it can be written as
\be
\Gamma^\mu(q^2) = g'\gamma^\mu F_1(q^2) + i \mu_\chi \sigma^{\mu\nu}q_\nu  ~F_2(q^2).
\label{eqn:Form-factors-def}
\ee
The first term is a charge-radius type operator and the neutrality of $\chi$ only demands that the associated form-factor vanishes at small momentum transfer $F_1(q^2) \rightarrow 0$ as $q^2\rightarrow0$. The second part of this vertex corresponds to an effective dipole operator for the WIMP as in Eq.~(\ref{eqn:MiDMInteraction}) with the dipole strength defined as \footnote{With this choice of the dipole moment, the absolute value of the one-loop form factor $F_2$ tends to one in the nonrelativistic, large messenger mass limit, see eq.~\eqref{eqn:F2nrlimit}.}
\be
\label{eqn:UVforMiDM}
\mu_\chi \equiv \frac{\lambda^2 g'}{32\pi^2\mmedf },
\ee
where $g'$ is the hypercharge coupling constant. Both form-factors were computed explicitly at one-loop in ref.~\cite{Weiner:2012gm} and the results are reproduced in Appendix~\ref{app:one-loop} for completion. 

Since this vertex corresponds to a non-renormalizable operator it is calculable since no local divergences arise at any order in perturbation theory. At one-loop level the apparent logarithmic divergence cancels precisely between the two diagrams of Fig.~\ref{fig:dipole_loop}. At two-loop level the leading (local) divergences should again cancel, but due to the appearance of sub-divergences we must include the existing counter-terms of the theory to arrive at a finite answer. This brings about the usual dependence of the final answer on the subtraction method and the renormalization scheme that was chosen to define the counter-terms. We therefore begin by defining the relevant counter-terms of the theory. We choose to work with dimensional regularization and minimal subtraction as the renormalization scheme, and we define
\be
d = 4 - 2 \epsilon, 
\ee
where $d$ is the dimension of spacetime. 

The fermion propagator counter-term at one-loop level is  associated only with corrections due to the Yukawa coupling, 
\be
\label{eqn:Fermion-propagator-ct}
\nonumber
i\mathcal{M}_{\rm FP}^{\rm (CT)}(p) &=& \parbox[t]{5cm}{\vspace{-1cm}
  \includegraphics[scale=0.6]{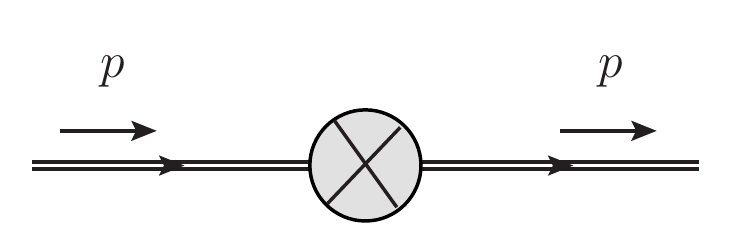} } \\  &=& -i\left(\tfrac{1}{2}\slashed{p} + \mX \right) \frac{\lambda^2}{16\pi^2} \MSdiv 
\\ \nonumber&\equiv& i\left(\delta_{Z\medf}\slashed{p} - \delta_{\mmedf} \right). 
\ee
The scalar propagator on the other hand receives corrections associated with the quartic interaction as well and is given by
\vspace{1cm}
\be
\label{eqn:Scalar-propagator-ct}
\nonumber
i\mathcal{M}_{\rm SP}^{\rm (CT)}(p)&=& \parbox[t]{5cm}{\vspace{-1.0cm}
  \includegraphics[scale=0.6]{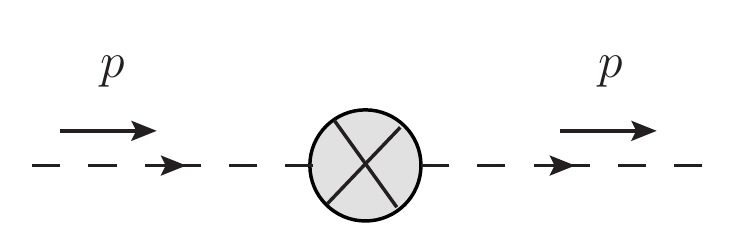} } 
\\ \nonumber &=& i \frac{\lambda^2}{8\pi^2}\times \left(-p^2 + 2(\mX^2+\mmedf^2+\mX \mmedf )\right) \MSdiv 
\\  &-& i \frac{\kappa}{16\pi^2} \left(\frac{N_{\meds}+1}{2}\times \mmeds^2\right)  \MSdiv 
\\ \nonumber 
&\equiv& i\left(\delta_{Z\meds}p^2 - \delta_{\mmeds} \right). 
\ee
Here $N_{\meds}$ is the dimension of the representation of $\meds$ under the electroweak $SU(2)$ group (e.g. $N_{\meds}= 2$ if $\meds$ is in the fundamental).  The mass and wave-function counter-terms for the fermion ($\delta_{\mmedf}$ and $\delta_{Z_{\medf}}$) and scalar ($\delta_{\mmeds}$ and $\delta_{Z_{\meds}}$) are defined through the above relations. Similarly we define the counter-terms for the interaction vertex of the fermion with the gauge field,
\vspace{1cm}
\be
\label{eqn:FFV-Vertex-ct}
\nonumber
i\mathcal{M}_{\rm FFV}^{\rm (CT)} \quad &=&\quad \parbox[t]{3cm}{\vspace{-1.5cm}
  \includegraphics[scale=0.6]{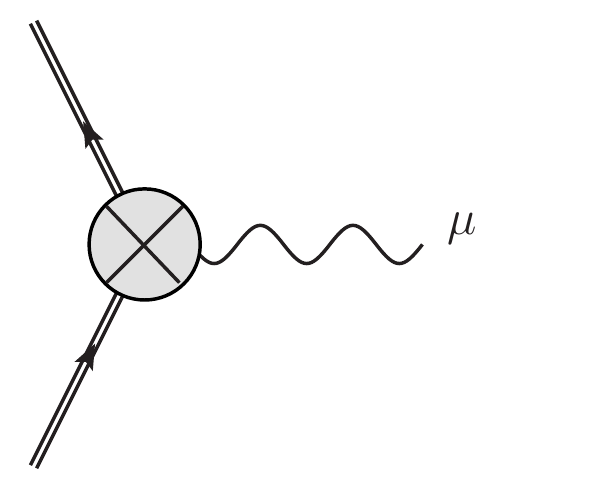} } 
\\ 
&=& -i g\gamma^\mu \frac{\lambda^2}{16\pi^2}\MSdiv \quad \equiv\quad ig\gamma^\mu \delta_{\rm FFV}, 
\ee
and the scalar with the gauge-field,
\vspace{1cm}
\be
\label{eqn:SSV-Vertex-ct}
\nonumber
i\mathcal{M}_{\rm SSV}^{\rm (CT)} \quad &=&\quad \parbox[t]{3cm}{\vspace{-1.5cm}
  \includegraphics[scale=0.6]{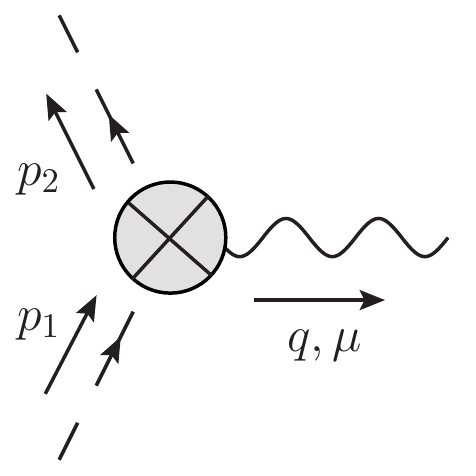} } 
\\   &=& i g \left(p_1^\mu+p_2^\mu \right) \frac{-(2+6y)\lambda^2}{16\pi^2} \MSdiv
\\ \nonumber
&\equiv& ig\left( p_1^\mu+ p_2^\mu\right)\delta_{\rm SSV}.
\ee

\begin{figure}
\begin{center}
\includegraphics[width=0.45 \textwidth]{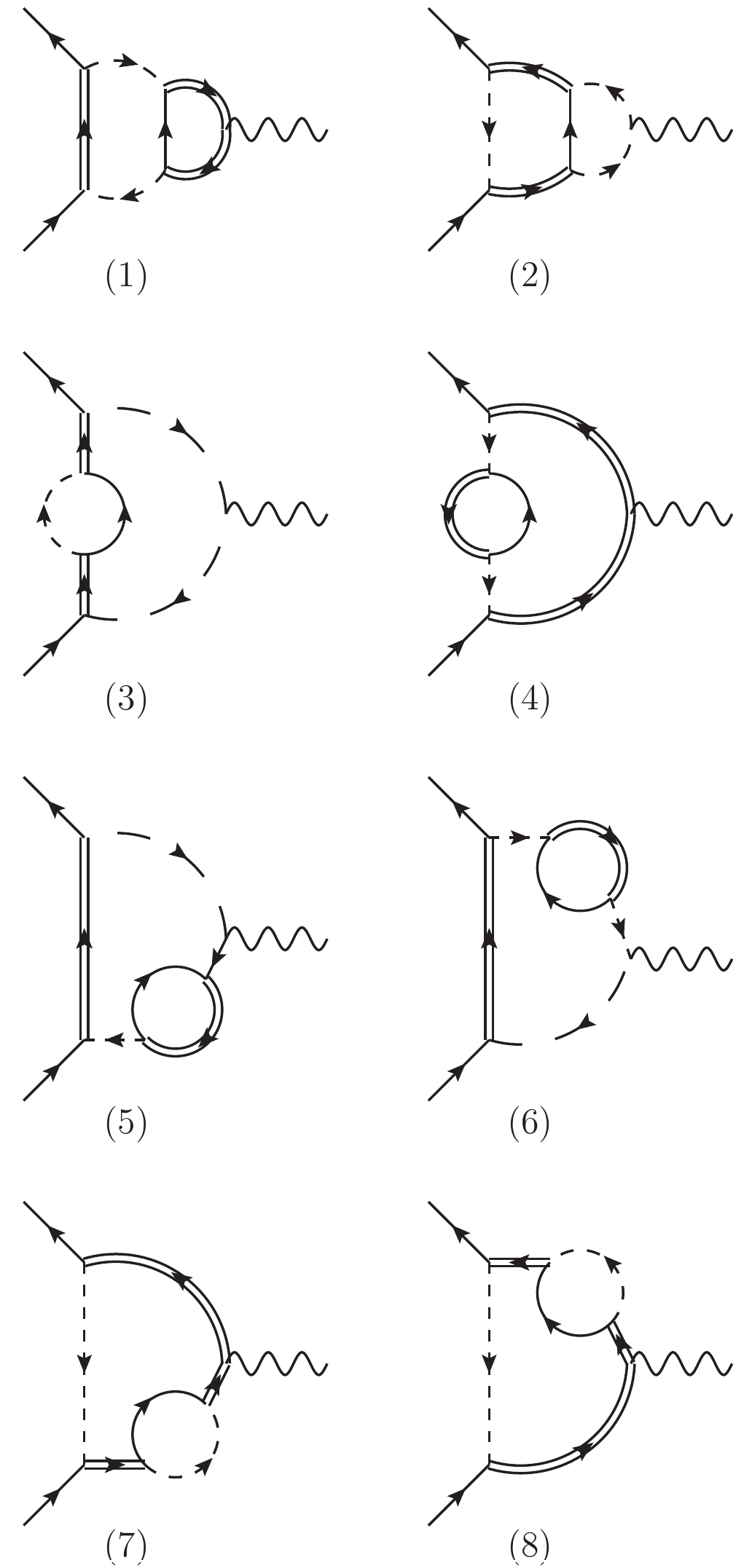}
\end{center}
\caption{All diagrams at two-loop order contributing to the magnetic dipole moment involving only the Yukawa interactions.}
\label{fig:all_2_loop_diagrams_Yukawa}
\end{figure}

In Fig.~\ref{fig:all_2_loop_diagrams_Yukawa} we depict all the diagrams at two-loop order which contribute to the magnetic dipole moment and involve only the Yukawa coupling. To each of these diagrams there exists a corresponding diagram with a counter-term inserted instead of the inner loop. This correspondence does not imply any mutual cancellation between the two contributions, but is simply a helpful organizational scheme. Only after all the diagrams, including the counter-terms, are added together do all the divergences cancel. The coefficients of the terms of order $\epsilon^{-2}$ are very simple and cancel separately among the diagrams in Fig.~\ref{fig:all_2_loop_diagrams_Yukawa} and among the diagrams involving the counter-terms. Using the same enumeration as in Fig.~\ref{fig:all_2_loop_diagrams_Yukawa} we find
\be
\epsilon^{-2}~{\rm coefficients} &=& i g \left(\frac{\lambda}{16\pi^2} \right)^2\frac{\gamma^\mu }{8}
\\ \nonumber &\times& \left(-4, 1, 1, -4, 4, 4, -1,  -1 \right).
\ee
The coefficients of the diagrams involving the counter-terms are obtained by multiplying the above by a factor of $(-2)$. Note that the sum vanishes as required by renormalizability of the theory and the neutrality of the WIMP. 

The coefficients of terms of order $\epsilon^{-1}$ correspond to non-local sub-divergences and are considerably more complex, involving all the masses in the problem. Generally these can only be written as unevaluated integrals over the Feynman parameters. Thus, their mutual cancellation forms a non-trivial check on the calculation. We have checked that they indeed cancel in two ways: by performing the integrals numerically for specific choices of masses and momenta; by considering the heavy mediator mass limit $\mmeds,\mmedf \gg \mX,q^2$, where the diagrams can be simplified. We now give the results for the latter case when the scalar and fermion mediators are set equal.
The terms involving $\sigma^{\mu\nu}q_\nu$ to order $\mmedf^{-1}$ in the heavy mediator expansion are given by 
\be
\label{eqn:epsilon_inverse_sigma_f_terms}
\left(
\begin{array}{c}
\epsilon^{-1}~{\rm coefficients} \\
\sigma^{\mu\nu}q_\nu~{\rm terms} 
\end{array}
\right)
&=& i g \left(\frac{\lambda}{16\pi^2} \right)^2 \frac{i \sigma^{\mu\nu}q_\nu}{48 \mmedf}
\\ \nonumber &\times& \left(-16,-8,4,40,20,20,6,6 \right),
\ee  
whereas the corresponding diagrams involving the counter-terms yield precisely the same results but with an opposite sign. Away from the equal mediator mass limit this sums up to
\be
&\sum_{\rm diag}& \left(
\begin{array}{c}
\epsilon^{-1}~{\rm coefficients} \\
\sigma^{\mu\nu}q_\nu~{\rm terms} 
\end{array}
\right) ~\\ \nonumber &=& i g \left(\frac{\lambda}{16\pi^2} \right)^2 \frac{2 i \mmedf \sigma^{\mu\nu}q_\nu}{\left( \mmedf^2-\mmeds^2 \right)^3}  \Bigg( -3 \mmedf^4 \\ \nonumber &+& 3\mmedf^2\mmeds^2 + \left(4\mmedf^4 +\mmedf^2\mmeds^2 + \mmeds^4 \right) \log\left(\frac{\mmedf}{\mmeds}\right)\Bigg),
\ee
which reproduces the sum of the terms in Eq.~(\ref{eqn:epsilon_inverse_sigma_f_terms}) in the appropriate limit. This contribution cancels exactly against the one coming from the diagrams involving the counter-terms. 

The terms involving $\gamma^\mu$ are given by
\be
\left(
\begin{array}{c}
\epsilon^{-1}~{\rm coefficients} \\
\gamma^{\mu}~{\rm terms} 
\end{array}
\right)
&=& i g \left(\frac{\lambda}{16\pi^2} \right)^2 \frac{\gamma^\mu}{48 \mmedf}
\\ \nonumber &\times& 
\left(
\begin{array}{c}
12+48\bar{\gamma} - 32\frac{\mX}{\mmedf} \\
15-12\bar{\gamma} + 8 \frac{\mX}{\mmedf} \\
-5-12\bar{\gamma} - 8 \frac{\mX}{\mmedf} \\
-68 + 48\bar{\gamma} - 112\frac{\mX}{\mmedf} \\
28 - 48\bar{\gamma} + 72\frac{\mX}{\mmedf} \\
28 - 48\bar{\gamma} + 72\frac{\mX}{\mmedf} \\
-5 +12 \bar{\gamma}  \\
-5 +12 \bar{\gamma} 
\end{array}
\right),
\ee 
which sums up to zero independently as can be confirmed directly. Here $\bar{\gamma} = \gamma - \log4\pi + \log \mmedf^2$. The corresponding terms involving the counter-terms are
\be
\left(
\begin{array}{c}
\epsilon^{-1}~{\rm coefficients} \\
\gamma^{\mu}~{\rm terms} 
\end{array}
\right)
&=& i g \left(\frac{\lambda}{16\pi^2} \right)^2 \frac{\gamma^\mu}{12 \mmedf}
\\ \nonumber &\times& 
\left(
\begin{array}{c}
-12\bar{\gamma} +8 \frac{\mX}{\mmedf} \\
3\bar{\gamma} -2 \frac{\mX}{\mmedf} \\
2+3\bar{\gamma} +2 \frac{\mX}{\mmedf} \\
8 - 12\bar{\gamma} +28 \frac{\mX}{\mmedf} \\
-4 +12\bar{\gamma} -18\frac{\mX}{\mmedf} \\
-4 +12\bar{\gamma} -18\frac{\mX}{\mmedf} \\
-1 -3 \bar{\gamma}  \\
-1 -3 \bar{\gamma} 
\end{array}
\right).
\ee

\begin{figure}
\begin{center}
\includegraphics[width=0.45 \textwidth]{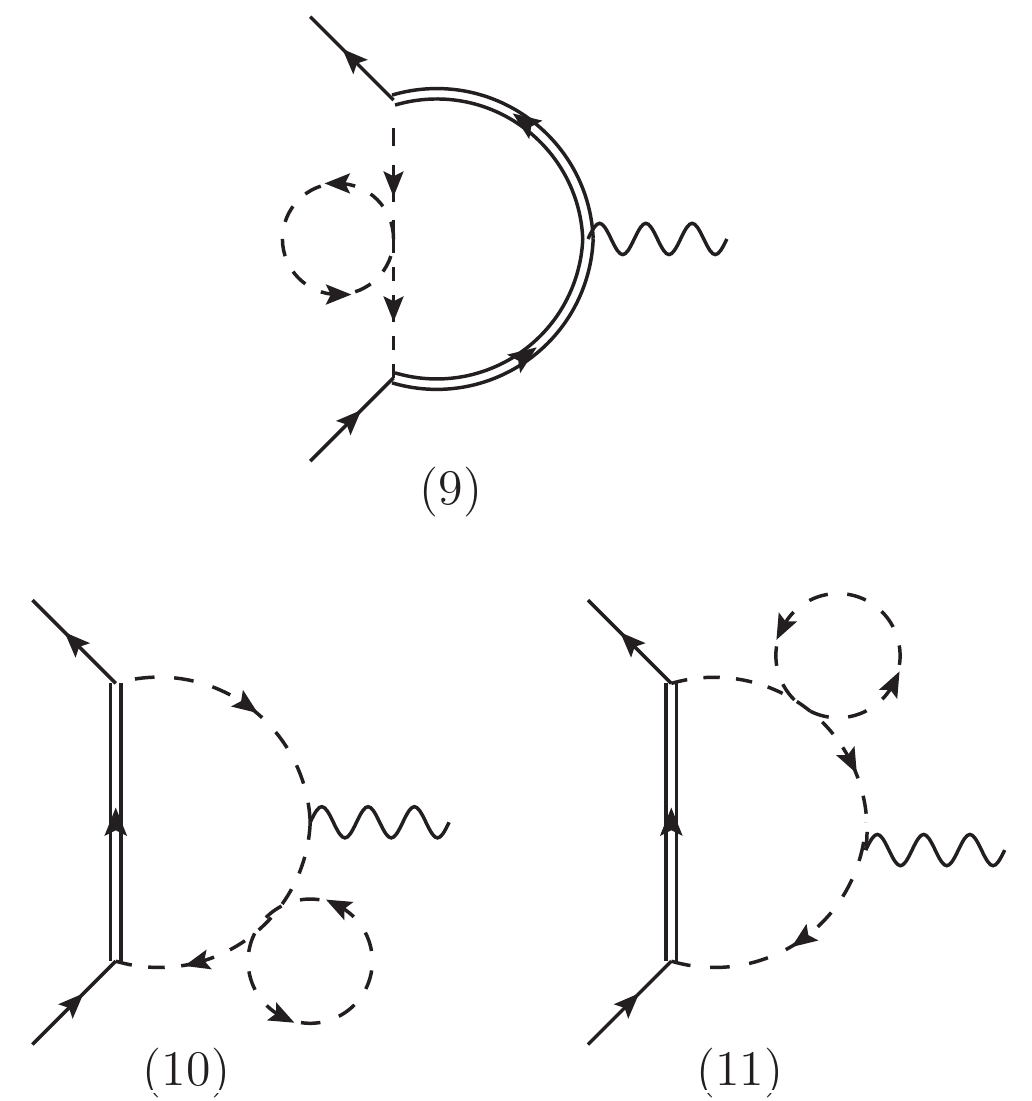}
\end{center}
\caption{All diagrams at two-loop orders contributing to the magnetic dipole moment involving the scalar quartic vertex.}
\label{fig:all_2_loop_diagrams_Quartic}
\end{figure}

If the quartic coupling $\kappa$ is included, then there are three more diagrams at two-loop order as shown in Fig.~\ref{fig:all_2_loop_diagrams_Quartic}.  At this order, there are no gauge-boson vertex corrections involving the quartic, only scalar propagator corrections. Since at one-loop order the quartic correction to the scalar propagator is independent of the momentum that flows into the loop, the associated two-loop diagrams are factorizable and hence easy to compute. The divergences in these diagrams are only of order $\epsilon^{-1}$ and cancel precisely against the contribution of the quartic to the counter-term of the scalar propagator in Eq.~(\ref{eqn:Scalar-propagator-ct}). We provide the full expressions associated with the diagrams in Fig~\ref{fig:all_2_loop_diagrams_Quartic} in Appendix~\ref{app:quartic_contribution}. 

  We calculated the diagrams in two different and independent ways: in the first approach we computed the diagrams analytically using Mathematica up to numerical integrals over Feynman parameters; in the second approach we used the Software \textsc{SecDec} \cite{Borowka:2012yc}; the results agreed throughout within the numerical uncertainties. \textsc{SecDec} can only compute Feynman integrals whose numerators are Lorentz scalars, so that the calculation of tensor integrals demands some additional work. In order to do so, one may start from the fact that, given the analyticity of the integrals in the dimensional regularization parameter $\epsilon$, and demanding Lorentz invariance, a tensor integral may be expressed as a Laurent series in $\epsilon$, where each term of the series is a sum of scalar functions of the external momenta, multiplied by monomials in the latter and the metric which are adequately contracted to reproduce the index structure of the integrand. Schematically, for a two-loop tensor integral $I_{2l}^{\mu_1,\dots,\mu_n}$ with external momenta $p_1,\dots p_k$,
  \begin{align}
   I_{2l}^{\mu_1,\dots,\mu_n}[p_1,\dots,p_k]=\sum_{n=-2}^\infty \epsilon^n \sum_m f_{n,m}(p_i\cdot p_j){\cal M}_{n,m}^{\mu_1,\dots,\mu_n},
  \end{align}
  where ${\cal M}_{n,m}^{\mu_1,\dots,\mu_n}$ represents a monomial constructed from the metric and the external momenta with the same index structure as the integrand, and the sum $m$ runs over all possible monomials. Symmetries of the integrand, for example under permutations of momenta, constrain the coefficients  $f_{n,m}(p_i\cdot p_j)$ and may be used to simplify the above expression. Furthermore, the functions $f_{-2,m}$ are restricted by the fact that in dimensional regularization the coefficient of the highest pole is a local polynomial of the external momenta of the appropriate mass dimension.  For example, an integral which is dimensionless in the limit $\epsilon\rightarrow0$ and which depends on a single external momentum $p$, can be written as
  \begin{align*}
   &\int d^Dk\, d^Dl\, l^\mu\, k^\nu F_{inv}(l,k,p)=\frac{1}{\epsilon^2}\,A_{-2} g^{\mu\nu}\\
   %%%%
   &+\frac{1}{\epsilon}(A_{-1}(p^2) g^{\mu\nu}+B_{-1}(p^2)p^\mu p^\nu)\\
   %%%%
   &+(A_0(p^2) g^{\mu\nu}+B_0(p^2)p^\mu p^\nu)+O(\epsilon).
  \end{align*}
  Note the absence of a $p^\mu p^\nu$ term in the $\epsilon^{-2}$ pole, which cannot appear in a local polynomial of the momenta with mass dimension zero.
  From the previous integral one may obtain scalar integrals by contracting the free Lorentz indices with the D-dimensional metric or with external momenta, which may be arbitrarily chosen \footnote{Note that \textsc{SecDec} assumes that the momenta of a diagram add up to zero, so that if one is to introduce additional spurious external momenta they have to add up to zero separately from the physical momenta.}. These scalar integrals can be computed with \textsc{SecDec}, and will depend as well on the $A$ and $B$ coefficient functions above; by using different contractions one obtains systems of equations that can be solved to obtain these coefficient functions evaluated at the chosen external momentum. The procedure can be generalized to more complicated Lorentz structures.

The results presented in the next section were obtained by numerically integrating over the Feynman parameters the analytic expressions for the integrands obtained in Mathematica. \textsc{SecDec} was used throughout to cross-check the real part of all the numerical results. It was also used to compute the imaginary part when kinematical thresholds were crossed (as in the calculation of the production cross-sections of the WIMP in colliders).

\section{Results}
\label{sec:results}

This section presents numerical results for the two-loop contributions for the form factors $F_1(q^2)$ and $F_2(q^2)$ in Eq.~\eqref{eqn:Form-factors-def}. This allows us to investigate the adequacy of the perturbation expansion in different regions of the parameter space. First, we will consider the case of time-like momentum exchange, which is relevant for WIMP annihilation as well as the direct production rate of WIMPs in colliders.  Second, we will treat the case of space-like exchange which is relevant for the phenomenology of DM direct-detection. Throughout we set the WIMP mass $\mX = 135\GeV$, but the results for any other value can be easily obtained by simple rescaling of all the mass scales involved (including the renormalization scale). 

\subsection{Time-like momentum exchange $q^2 > 0$}
\label{subsec:timelike}

\begin{figure}
\begin{center}
\includegraphics[width=0.45 \textwidth]{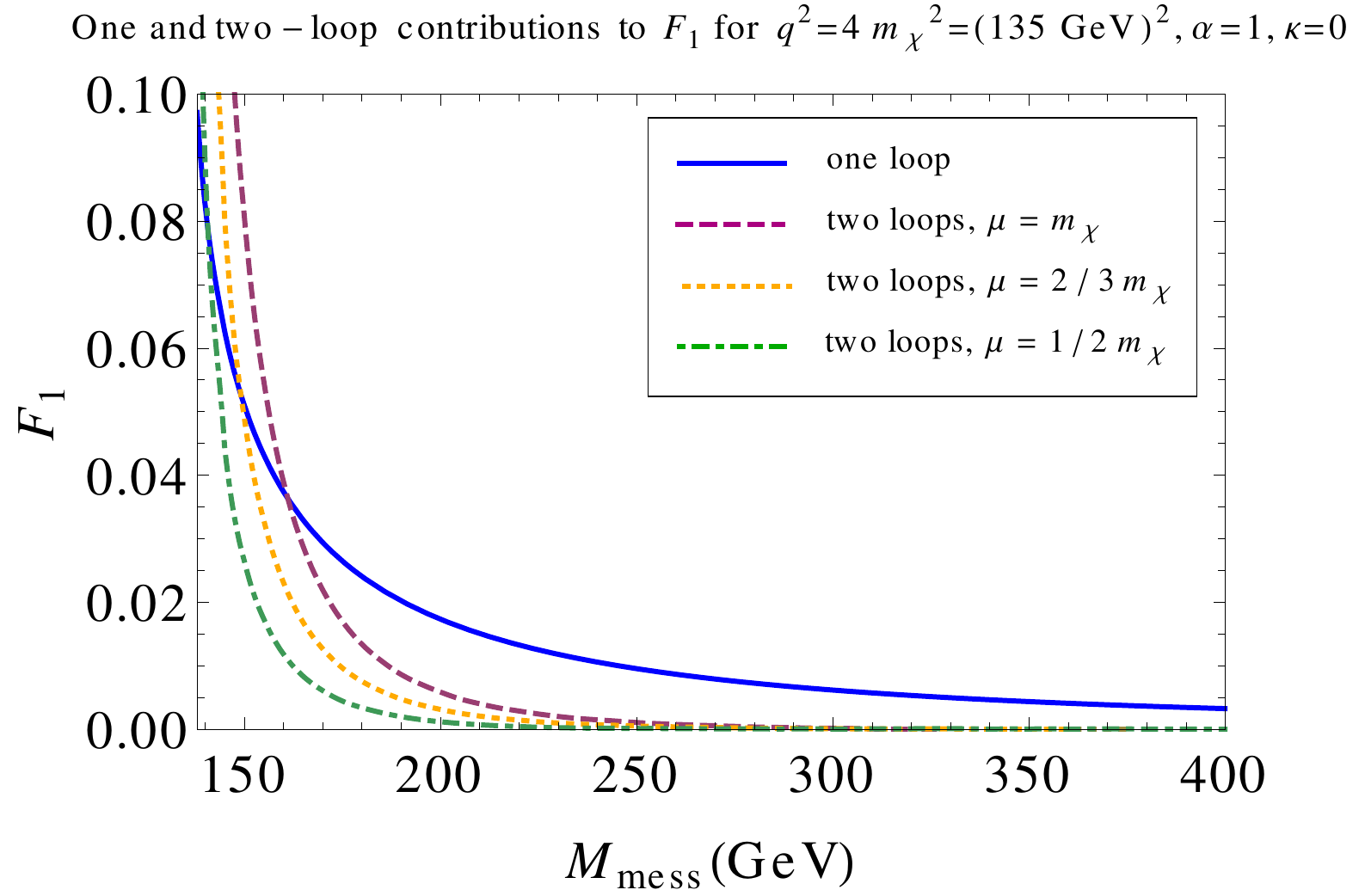}
\vskip.4cm
\includegraphics[width=0.45 \textwidth]{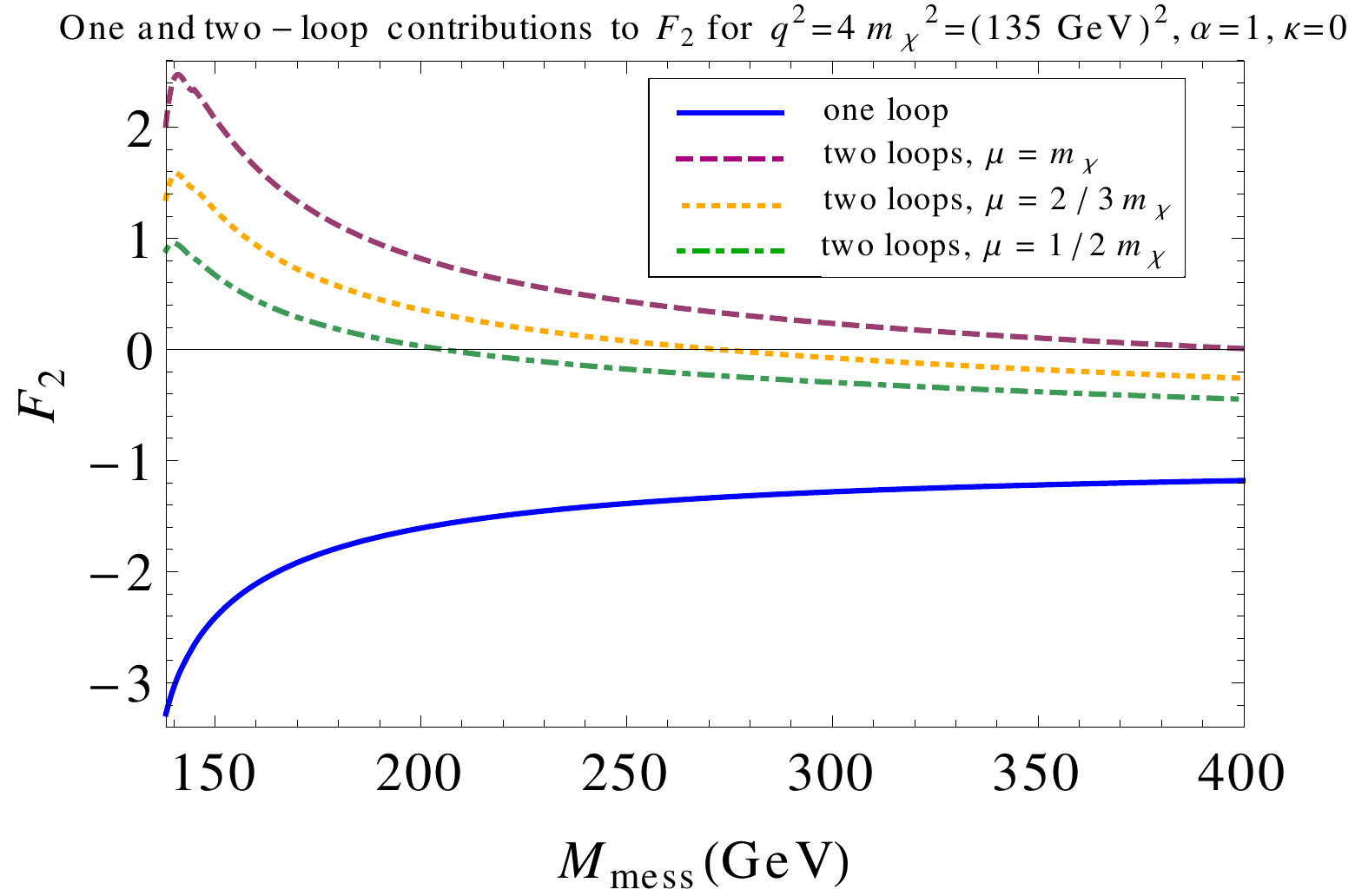}
\vskip.4cm
\end{center}
\caption{One- and two-loop contributions to the form factors $F_1$ (top-pane) and $F_2$ (bottom-pane) as a function of the messenger mass, for different values of the RG scale. Here we set the WIMP mass $\mX=135$ GeV, the messenger masses $\mmeds = \mmedf = M_{mess}$, the momentum transfer $q^2=4 \mX^2$, and the couplings $\alpha_\lambda=1$, $\kappa=0$.}
\label{fig:F1_timelike}
\end{figure}

A time-like momentum exchange $q^2=(2\mX)^2 = 4 \mX^2$ is relevant for annihilation calculations in the non-relativistic limit. In ref.~\cite{Weiner:2012gm} it was shown that this process leads to the correct relic abundance of the WIMP when the coupling to the charge messengers is sizable $\alpha_\lambda \approx 1-2$ for messenger masses in the range of $140 - 300\GeV$. In Fig.~\ref{fig:F1_timelike} we show numerical results for the one and two-loop
contributions to the form factors $F_1$ and $F_2$ coming from the diagrams of section \ref{sec:Setup} for different values of the renormalization scale $\mu$. Note that $\mu$ was changed while keeping the Yukawa coupling $\lambda$ constant, so that we were not probing the total renormalization group (RG) dependence, but rather the explicit dependence on $\mu$. As such, the one-loop correction has no explicit dependence on $\mu$ since it is finite as dictated by gauge invariance (dark matter is neutral and has no tree-level couplings to the photon). This is in contrast with the two-loop corrections, which do explicitly depend on $\mu$ due to the presence of one-loop sub-divergences. This is expected from the fact that the resummed Green function has to be independent of $\mu$, so that the implicit dependence on the scale through the renormalization group has to be compensated by an explicit dependence. Now, since the resummed Green function is scale-independent, one is free to choose the value of $\mu$ at one's convenience, but when truncating the perturbative expansion it makes sense to choose the value of the scale that 
minimizes the higher order corrections and leads to a better perturbative behavior. From Fig.~\ref{fig:F1_timelike} we see that even for a strong value of the coupling such as $\alpha_\lambda=1$, two loop corrections can be kept smaller than the one-loop ones by more than a factor of two for a wide range of messenger masses. This range can be optimized by choosing an appropriate renormalization scale; for $M_{\rm mess} \lesssim 200\GeV$ for example a good choice is $\mu \approx \mX/2$ as can be seen in Fig.~\ref{fig:F1_timelike}. This is suggestive of a reliable perturbative behavior, so that the results of ref.~\cite{Weiner:2012gm} with the two-loop update presented here may be trusted when calculations are performed at an adequate RG scale. Results for other values of the coupling $\lambda$ can be obtained by taking into account that the one and two-loop contributions to the form-factor $F_1$ scale as $\lambda^2$ and $\lambda^4$, respectively. The results  for the magnetic form-factor $F_2$ scale as a constant (one-loop) and $\lambda^2$ (two-loop). Even for larger values of the coupling $\alpha_\lambda > 1$, an acceptable perturbative behavior may be maintained for sufficiently large messenger masses.

\begin{figure}
\begin{center}
\includegraphics[width=0.45 \textwidth]{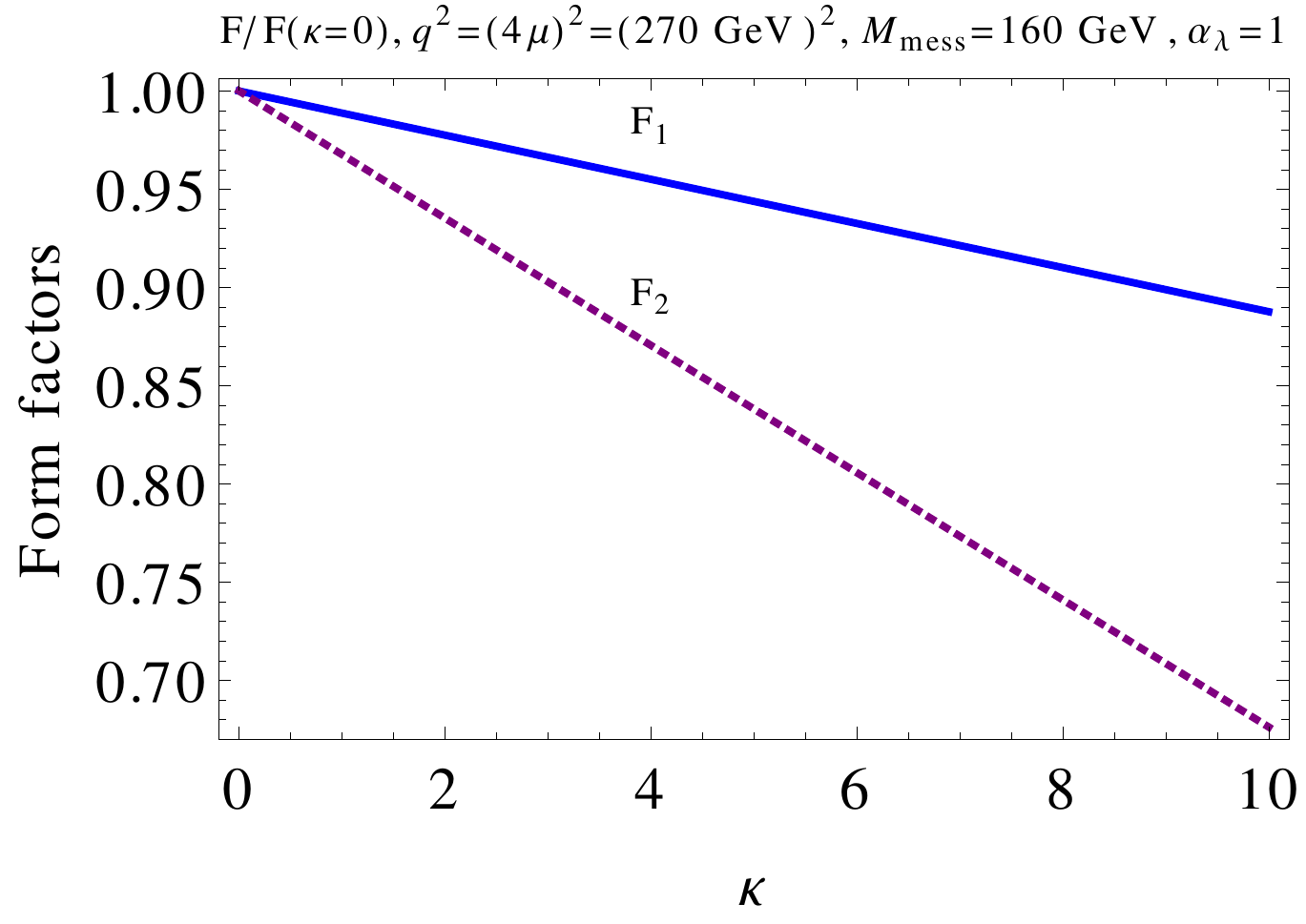}
\end{center}
\caption{Two-loop contributions to the form factors $F_1$ and $F_2$ as a function of the quartic coupling $\kappa$. The form-factors are normalized to their value at $\kappa=0$. Here we set the WIMP mass $\mX=2\mu=135$ GeV, the messanger mass $\mmedf=\mmeds=160$ GeV, and the Yukawa coupling $\alpha_\lambda=1$.
%% F1: 1. - 0.0112326 x,   F2:1. - 0.0323749 x
}
\label{fig:Fskappa}
\end{figure}

Regarding the dependence on the quartic coupling $\kappa$, the diagrams of Fig.~\ref{fig:all_2_loop_diagrams_Quartic} give rise to contributions proportional to $\kappa\lambda^2$, with coefficients that turn out to be smaller than those of the $\lambda^4$ contributions. These are subdominant as long as $\kappa\lesssim\lambda^2$.  Whether or not this is the case cannot be presently determined since in contrast with the Yukawa coupling, there is no direct phenomenological constraint on the quartic ($\kappa$ only affects dark-matter annihilation diagrams at two-loops and beyond, so that the one-loop results of
ref.~\cite{Weiner:2012gm} only enforce $\lambda$ to be large). Nevertheless it is reassuring that the phenomenology of the model is not strongly sensitive to the choice of the quartic as seen in Fig.~\ref{fig:Fskappa} which shows the effect of the quartic $\kappa$ on the form-factors.

Values of the momentum transfer greater than $4 m^2_\chi$ are relevant for the estimation of cross-sections at the LHC, which are studied in \S \ref{subsec:LHCprod}. The appearance of threshold effects and numerical stability makes a full computation of the two-loop diagrams for an appreciable range of $q^2$ challenging. The program \textsc{SecDec} can successfully deal with threshold singularities, yet numerical issues are still problematic. As explained in \S \ref{subsec:LHCprod}, we were able to compute the one-loop form factors, including their imaginary parts, for the whole range of values of $q^2$ relevant for LHC phenomenology (see Fig.~\ref{fig:Fs_imaginary}). As will be argued in \S~\ref{subsec:LHCprod}, the LHC cross-sections are expected to be dominated by $F_2$ evaluated near thresholds, where the one-loop contribution is greatly enhanced with respect to the two-loop one. This can be seen on the bottom pane of Fig.~\ref{fig:F1_timelike} near the threshold for messenger pair production, i.e. for  $M_{mess}\sim m_\chi$ (recall that for this figure  $q^2=4 m_\chi^2$). Therefore, while we do not provide results for the two-loop contributions in the region $q^2 > 4\mX^2$, we believe these effects to be inconsequential to LHC phenomenology. 

\subsection{Space-like momentum exchange $q^2 < 0$}

The scattering process of WIMPs against nuclei and the subsequent observable nuclear recoil forms the basis for efforts of direct-detection of DM in underground laboratories. Since the WIMP is entirely non-relativistic and the target is at rest, these collisions are dominated by very low space-like momentum exchange $q^2 \ll \mX$. Thus, it is useful to expand the form-factors in powers of the momentum exchange and keep only the leading order terms. Since the WIMP is neutral, $F_1$ must vanish in the limit of $q^2\rightarrow 0$ and so it proves useful to define the following,
\be
\label{eqn:lowQ_F1}
F_1(q^2) = -\frac{\muX q^2}{6g'\mmedf} f_1.
\ee
In the limit of small momentum-transfer, $q^2\rightarrow 0$, the form-factors at one-loop are then given by
\be
\label{eqn:f1_1_loop}
\nonumber f_1^{\rm (1)} &=&- \frac{3}{\mOverM^2}\left(1-\frac{\sqrt{1-\mOverM^2}}{\mOverM} \tan^{-1}\left( \frac{\mOverM}{ \sqrt{1-\mOverM^2} }\right)\right) \\
&=&- 1- \frac{2\mOverM^2}{5} + \mathcal{O}\left( \mOverM^4 \right), \\
\nonumber 
\\\nonumber
\label{eqn:F2_1_loop}
F_2^{\rm (1)} &=& -\frac{1}{\mOverM}\left(1-\frac{1}{\mOverM} \sqrt{\frac{1+\mOverM}{1-\mOverM}}\tan^{-1}\left( \frac{\mOverM}{ \sqrt{1-\mOverM^2} }\right)\right) \\ 
&=& -1- \frac{2\mOverM}{3}- \frac{2\mOverM^2}{3} + \mathcal{O}\left( \mOverM^3 \right),
\ee
where
\be
\mOverM \equiv \frac{\mX}{2\mmedf}.
\ee
Defined in this way, both $f_1$ and $F_2$ approach unity in the heavy mediator mass limit $\mmedf \gg \mX$.

We begin by giving the two-loop contribution to the form-factor coming from the quartic coupling since it can be stated analytically,
%\onecolumngrid
%\begin{equation}
\begin{flalign}
\nonumber 
~& \Big[\frac{\kappa}{16\pi^2}\left(\gamma-1 + \log\left(\frac{\mmeds^2}{4\pi \mu^2} \right)\right)\Big]^{-1} f_1^{(2)}  =  \\ \nonumber
~&\frac{3\left(\mOverM  \sqrt{1\!-\!\mOverM ^2} \left(\!-3\!+\!2 \mOverM ^2\right)\!+\left(3\!-\!4 \mOverM ^2\right) \tan^{-1}\left[\frac{\mOverM }{\sqrt{1-\mOverM ^2}}\right]\right)}{8(1-\mOverM )\mOverM ^4 \sqrt{1-\mOverM ^2}}
\\
~&=-\frac{2\mOverM}{5} -\frac{2\mOverM^2}{5}+\mathcal{O}\left( \mOverM^3 \right),~
\end{flalign}
%\begin{equation}

\begin{flalign}
\nonumber
~&\Big[\frac{\kappa}{16\pi^2}\left(\gamma-1 + \log\left(\frac{\mmeds^2}{4\pi \mu^2} \right)\right)\Big]^{-1} F_2^{(2)}  = \\ \nonumber
~&\frac{3 \left(\mOverM  \sqrt{1-\mOverM ^2}+\left(-1+2 \mOverM ^2\right) \text{ArcTan}\left[\frac{\mOverM }{\sqrt{1-\mOverM ^2}}\right]\right) }{8(1-\mOverM )\mOverM ^3 \sqrt{1-\mOverM ^2}}
\\
~&= \frac{1}{2}+\frac{\mOverM }{2}+\frac{4 \mOverM^2}{5}+\mathcal{O}\left( \mOverM^3 \right).
\end{flalign} 
%\twocolumngrid
Unless the quartic approaches non-perturbative values, $\kappa \approx 16\pi^2$, these contributions remain small as compared with the one-loop results.

\begin{figure}
\begin{center}
\includegraphics[width=0.45 \textwidth]{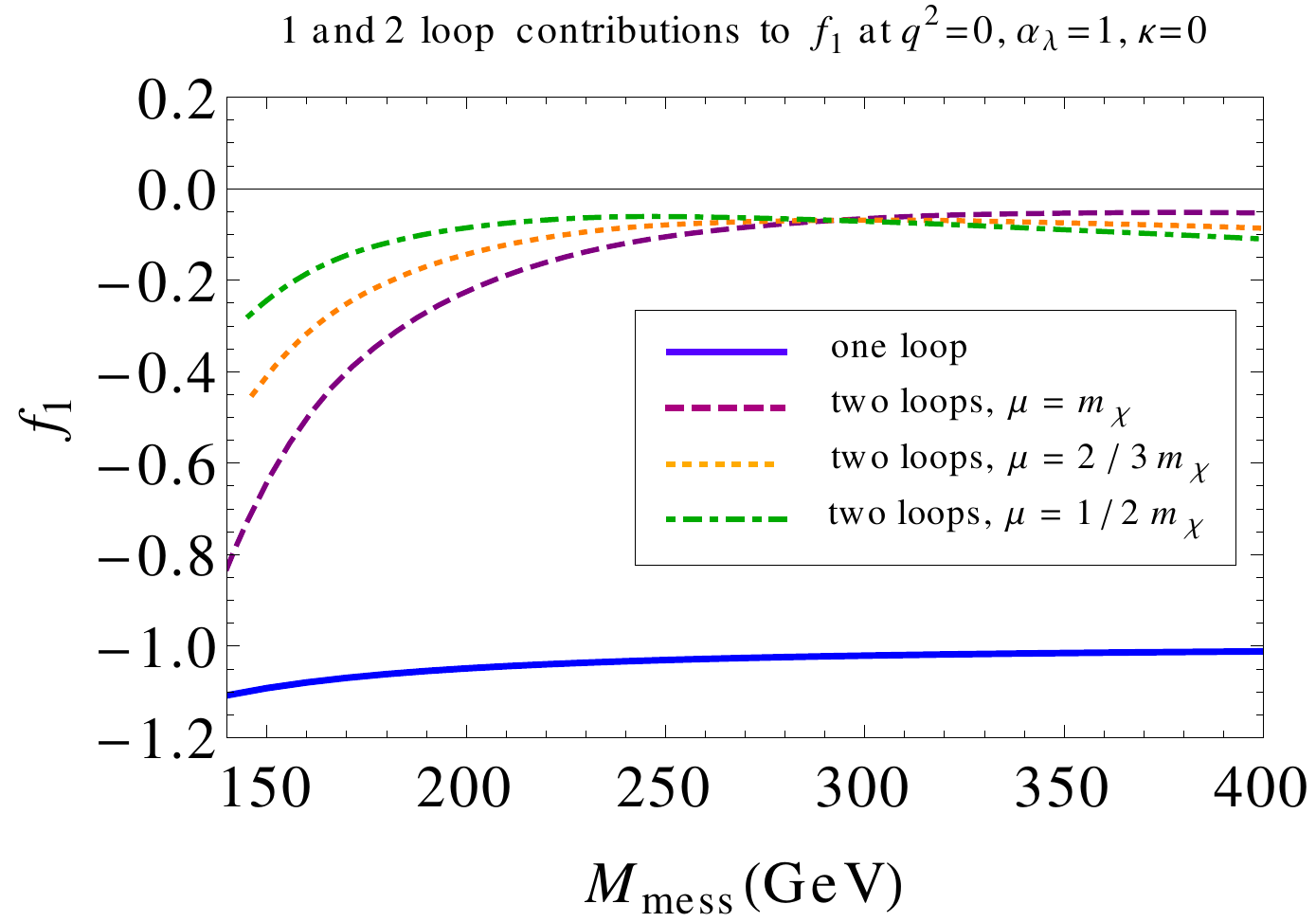}
\vskip.2cm
\includegraphics[width=0.45 \textwidth]{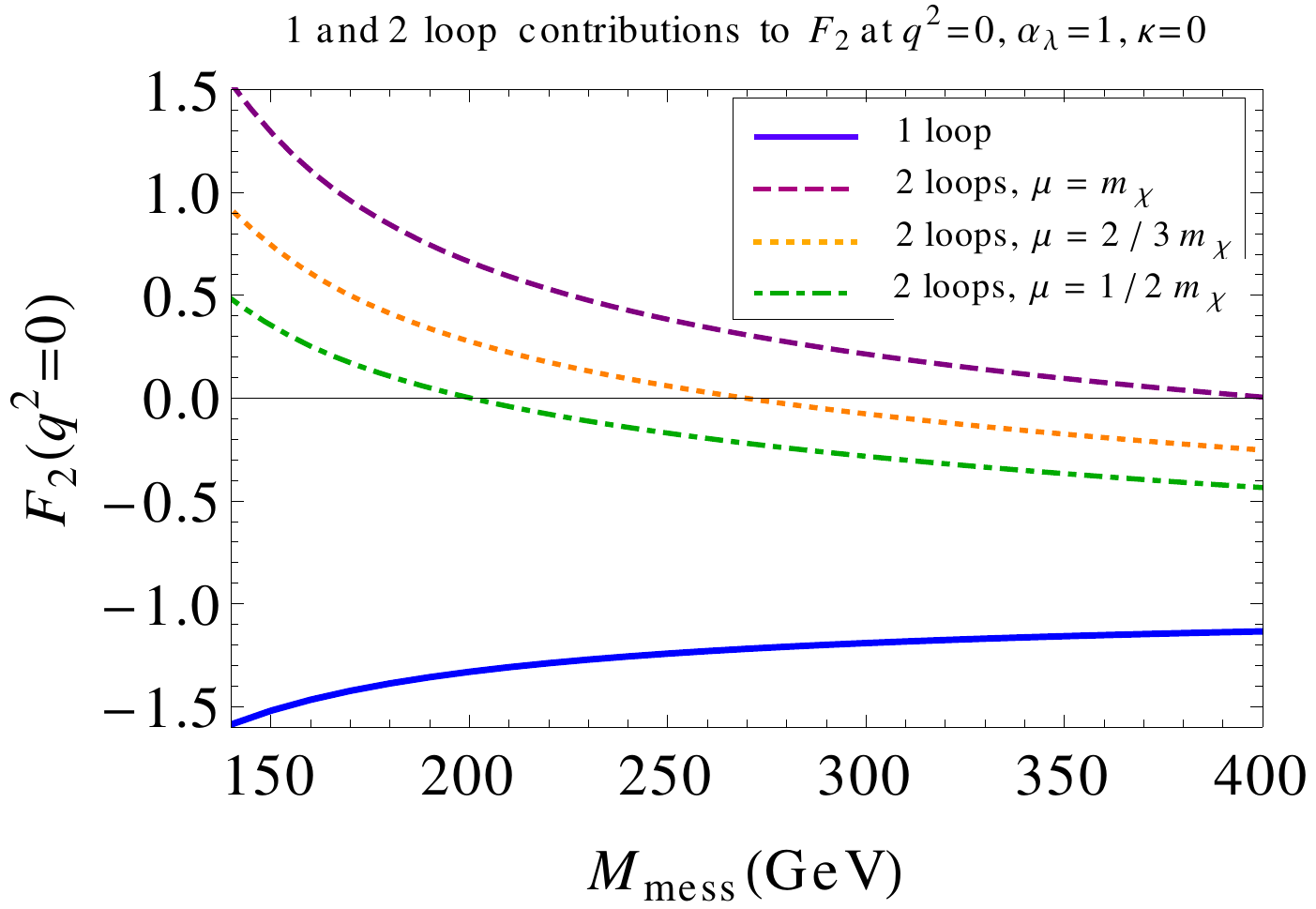}
\end{center}
\caption{One- and two-loop contributions to the form factors $f_1$ (up) and $F_2$ (down) in the low momentum exchange limit $q^2=0$. Here $\mX=135$ GeV,  $\alpha_\lambda=1$, $\kappa=0$, in terms of the messenger mass, with one set of messenger fields, for different values of the renormalization scale $\mu$.}
\label{fig:FF_lowQ}
\end{figure}

The two-loop corrections associated with the Yukawa coupling alone are more involved and cannot be easily written down in closed form.  Instead we plot the form-factors as a function of the messenger mass in Fig.~\ref{fig:FF_lowQ}. Again the two-loop contribution is more substantial for very low messenger masses $\mmedf,\mmeds \approx \mX$, but remains under control  for the appropriate choice of renormalization scale $\mu \approx m_\chi/2$.

\section{Phenomenology}
\label{sec:pheno}

\subsection{Relic Abundance}
\label{subsec:relic_abundance}

Armed with the numerical results summarized in the previous section, we can explore the phenomenological consequences. The most immediate question to be answered is how the two-loop corrections affect the value of the coupling required to get the correct relic abundance --indeed, it was the requirement of getting the relic-abundance correct at one-loop order that demanded a strong  coupling $\alpha_\lambda \approx 1$ in the first place \cite{Weiner:2012gm}. The relic abundance can be computed from the annihilation cross-sections into light fermions and gauge bosons, which are 
given in appendix \ref{app:cross_section_formulae}. Fig.~\ref{fig:alphaVsMmed} shows the value of the coupling $\alpha_\lambda$ that yields the correct relic abundance as a function of a common messenger mass, for different values of the RG scale, at one and two-loops.
As noted in the previous section, two-loop corrections are under control and do not drastically change the phenomenology.  Again, these corrections can be minimized for appropriate choices of the renormalization scale; as was seen before, for low messenger masses an optimal choice is $\mu \approx \mX/2$, for which the correct relic abundance can be obtained while keeping a reasonable perturbative behavior,
with two-loop contributions at least a factor of two smaller than the one-loop result. This is 
shown in more detail in Fig.~\ref{fig:Fratios}, where we plot the ratios between the one- and two-loop contributions to the form factors $F_1$ and $F_2$ when $\alpha_\lambda$ is fixed to get 
the correct relic abundance and the RG scale is set at $\mu =\mX/2$. 

It can also be seen in Fig.~\ref{fig:alphaVsMmed} that both the one and two-loop results converge towards similar small values of these coupling for small enough messenger masses. The annihilation cross-section is dominated by the contributions from the form factor $F_2$, as evidenced by their numerical values in Fig.~\ref{fig:F1_timelike} and the fact that the zeros of the two-loop contributions to $F_2$ of Fig.~\ref{fig:F1_timelike} correspond quite closely to the crossings of the one- and two-loop curves of Fig.~\ref{fig:alphaVsMmed}. When the messenger mass approaches the dark matter  mass the momentum transfer in the non-relativistic limit $q^2=(2 \mX)^2$ is close to the threshold singularity associated with the on-shell production of a pair of messengers, so that the annihilation cross-section is enhanced and a lower value of $\alpha_\lambda$ is required to get the relic abundance. The fact that the one and two-loop curves converge can be explained by noting that for $F_2$ near the threshold the two-loop results remain bounded, while the one-loop result grows fast in magnitude and dominates (see Fig.~\ref{fig:F1_timelike}).

It is interesting to note that for larger representations of the messenger fields, the contributions from the one- and two-loop diagrams both scale as the number of messengers in the loop (neglecting the quartic). Therefore, at least at this order of perturbation theory, it is possible to obtain the correct relic abundance with a lower interaction strength, but larger representations. This has the effect of diminishing the relative importance of the two-loop contribution, as is clearly illustrated by the bottom pane of Fig.~\ref{fig:alphaVsMmed} in the case of two families of messengers.

\begin{figure}
\begin{center}
\includegraphics[width=0.45 \textwidth]{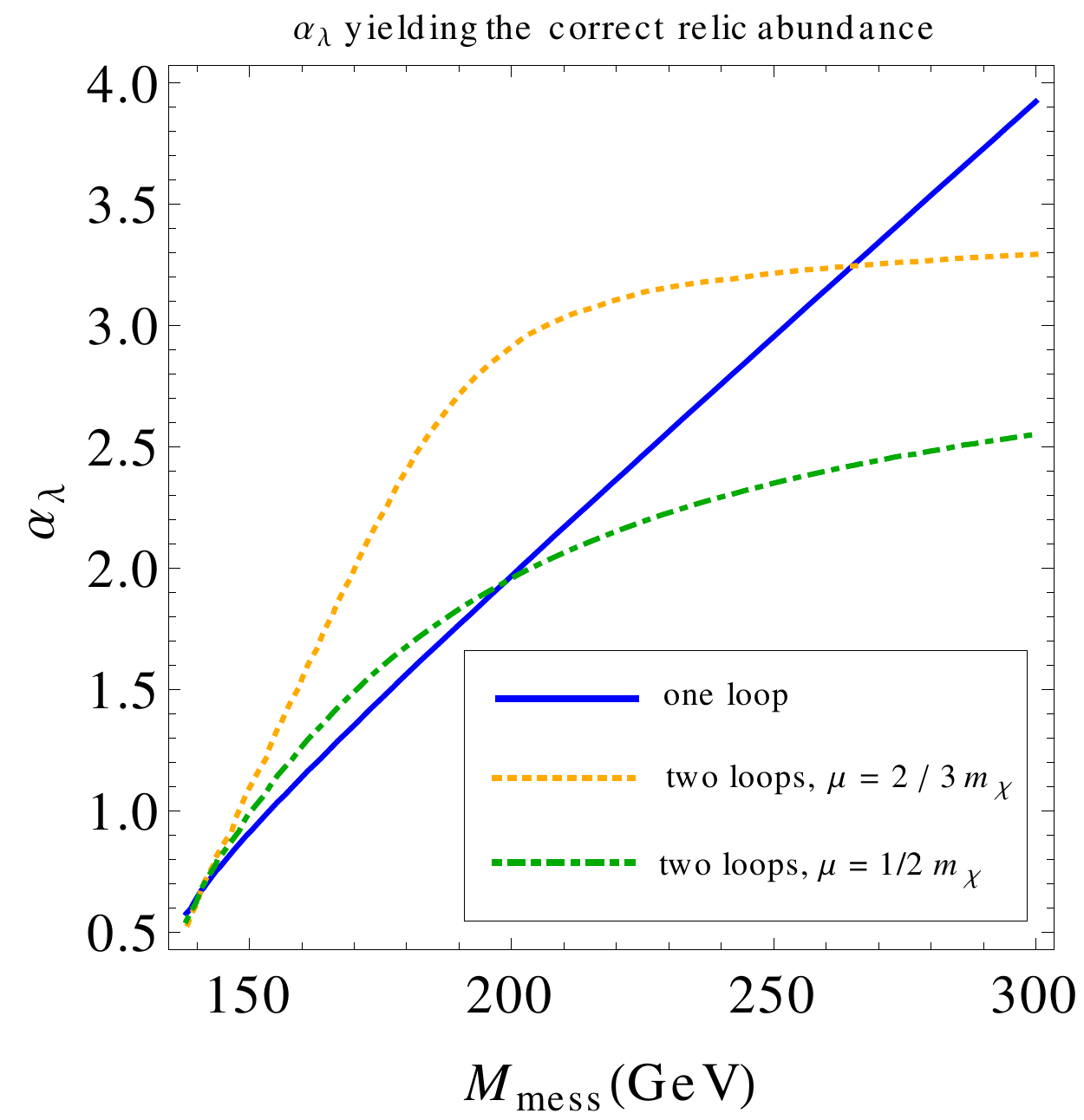}
\includegraphics[width=0.45 \textwidth]{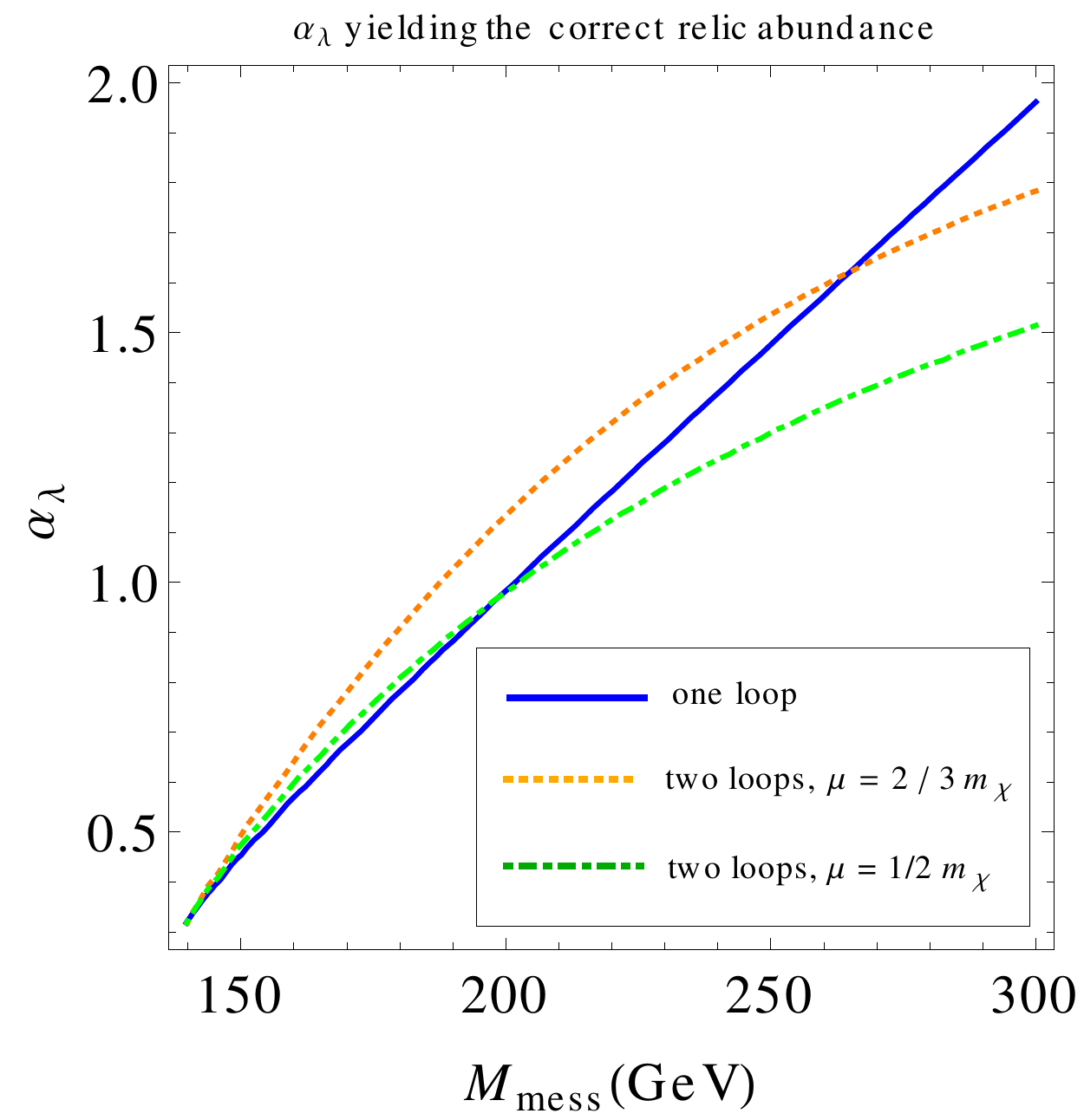}
\end{center}
\caption{The Yukawa coupling vs. the messengers' mass fixing the freezeout rate at $\sigma v = 6\times 10^{-26}{\rm cm^3/s}$, at one and two-loop order, for different values of the RG scale $\mu$, in the case of one family of messengers (top pane) and two families (bottom pane). Note that the two-loop corrections  in the latter case are relatively smaller, as discussed in the text.}
\label{fig:alphaVsMmed}
\end{figure}

\begin{figure}
\begin{center}
\includegraphics[width=0.45 \textwidth]{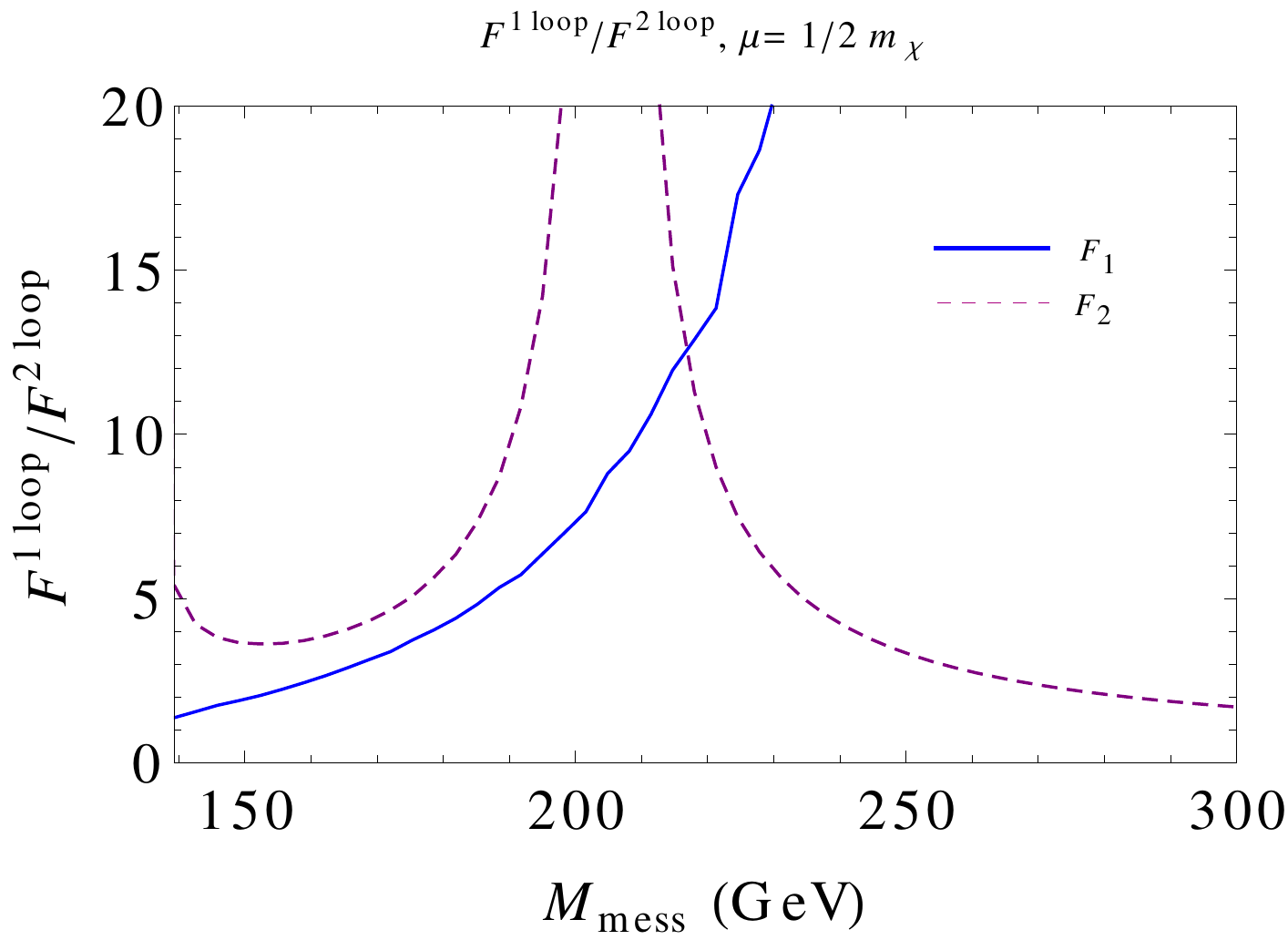}
\end{center}
\caption{Ratios between the one and two-loop contributions to the form factors $F_1$ and $F_2$, for $\mX=2\mu=135$ GeV, $q^2=4 \mX^2$,  $\kappa=0$, and $\alpha_\lambda$ fixed by 
the relic abundance as in Fig.~\ref{fig:alphaVsMmed}, for one messenger family. The jagedness of the solid curve is due to numerical precision issues coming from the fact that the overall form factor $F_1$ is close to zero (see Fig.~\ref{fig:F1_timelike}). In the case of $F_2$, the ratio of one-to two-loop contributions blows up as the two-loop correction approaches zero while the one-loop result stays finite, as can be seen from Fig.~\ref{fig:F1_timelike}.}
\label{fig:Fratios}
\end{figure}

The results show that the two-loop contributions to the annihilation cross section are under control, and a correct relic abundance can be still obtained. This boosts our confidence in the fact that the phenomenological success of the model is not foiled by higher loop corrections and is robust despite the strong coupling. However, it should be kept in mind that the above only pertains to the s-channel annihilation into charged fermions (and vector-bosons) through $\gamma/\ZZ$ which determines the relic abundance of DM in this model. The annihilation into two photons $\gamma\gamma$ (or $\gamma+\ZZ$), which is relevant for the rate observed in searches for $\gamma$-ray lines, is dominated by the Rayleigh operator~\cite{Weiner:2012cb,Weiner:2012gm}. In the present work we have not computed the two-loop contribution to the Rayleigh operator and so it is not yet possible to determine the effect on the rate at two-loop order. 

\subsection{Direct Detection}

We now turn to another aspect of the phenomenology of MDM, namely the possibility of directly detecting DM collisions with  nuclei by scattering through its dipole moment. The relevant phenomenological observable is the nuclear recoil rate and this has been calculated for elastic MDM models~\cite{Barger:2010gv,Banks:2010eh} as well as the inelastic MiDM model~\cite{Chang:2010en}. For scattering against nuclei photon exchange is dominant and the relevant vertex is as defined in Eq.~(\ref{eqn:Form-factors-def}), but with $\muXg \equiv \muX \cos\theta_{_W}$. At low momentum exchange it then proves useful to use Eq.~(\ref{eqn:lowQ_F1}), $F_1(q^2) = -\frac{\muXg q^2}{6g'\mmedf} f_1$ and write the vertex as
\be
\label{eqn:lowQ_vertex}
\Gamma^\mu(q^2) = -\gamma^\mu \left( \frac{ \muXg q^2}{6\mmedf} \right)f_1+ i \muXg \sigma^{\mu\nu}q_\nu  ~F_2.
\ee 
The second part of the vertex associated with the dipole moment leads to scattering against the nucleus' dipole as well as its charge. We consider then the scattering process $\chi N \rightarrow \chi' N$ where $N$ represents the nucleus, and $\chi'$ is some possibly heavier WIMP states with a mass splitting $\mX + \delta$. The elastic limit can be easily obtained from the results below by setting $\delta = 0$. Ignoring the contribution from $f_1$ for a moment, the differential cross-section for non-relativistic  inelastic WIMP scattering against the nucleus due to the magnetic dipole moment can be written as~\cite{Chang:2010en}
\be
\label{eqn:QEDformula}
\frac{d \sigma}{d \ER} = \frac{d\sDD}{d \ER}+\frac{d \sDZ}{d \ER},
\ee
where the dipole-dipole term is given by,
\begin{eqnarray}
\label{eqn:sigmaDD}
\nonumber
\frac{d \sDD}{d \ER}&=& Z_\chi\frac{16 \pi \alpha^2  \mN}{v^2} \left(\frac{\mu_{nuc}}{e}\right)^2 \left(\frac{\muXg}{e}\right)^2 \\[-.2cm]
\\[-.2cm]
\nonumber
 &\times& \left(\frac{S_\chi+1}{3 S_\chi}\right) \left(\frac{S_N+1}{3 S_N }\right)F_D^2[\ER],
\end{eqnarray}
and the dipole-charge term is given by,
\begin{eqnarray}
\label{eqn:sigmaDZ}
\nonumber
\left.\frac{d \sDZ}{d \ER}\right|_{F_1=0}&=& Z_\chi\frac{4\pi Z^2 \alpha^2}{\ER}\left(\frac{\muXg}{e}\right)^2 \bigg[1\!-\!\frac{E_R}{v^2}\left(\frac{1}{2m_N}\!+\!\frac{1}{\mX}\!\right)\\[-.12cm]
\\[-.15cm]
\nonumber
&-&\frac{\delta}{v^2}\left(\frac{1}{\mu_{\scriptstyle{N\chi}}}+\frac{\delta}{2m_N E_R}\right)\bigg]
\left(\frac{S_\chi+1}{3S_\chi}\right) F^2[\ER].
\end{eqnarray}
Here $\ER$ is the recoil energy of the nucleus as measured in the lab, $\mN$ is the mass of the nucleus, $\mu_{nuc}$ is the nuclear magnetic dipole moment, $S_\chi = 1/2$ is the spin of the WIMP and $S_N$ is the spin of the nucleus, $F_D[\ER]$ is the nuclear spin form-factor, $Z$ is the nuclear charge, and $F[\ER]$ is the nuclear charge form-factor. We included the dark-matter field renormalization factor $Z_\chi$ that enters the scattering amplitude at two-loops and beyond (see Eqs.~\eqref{eqn:ZDM2} and \eqref{eqn:ZDM}). A detailed discussion of these terms and their affects can be found in ref.~\cite{Chang:2010en}.

\begin{figure}
\begin{center}
\includegraphics[width=0.45 \textwidth]{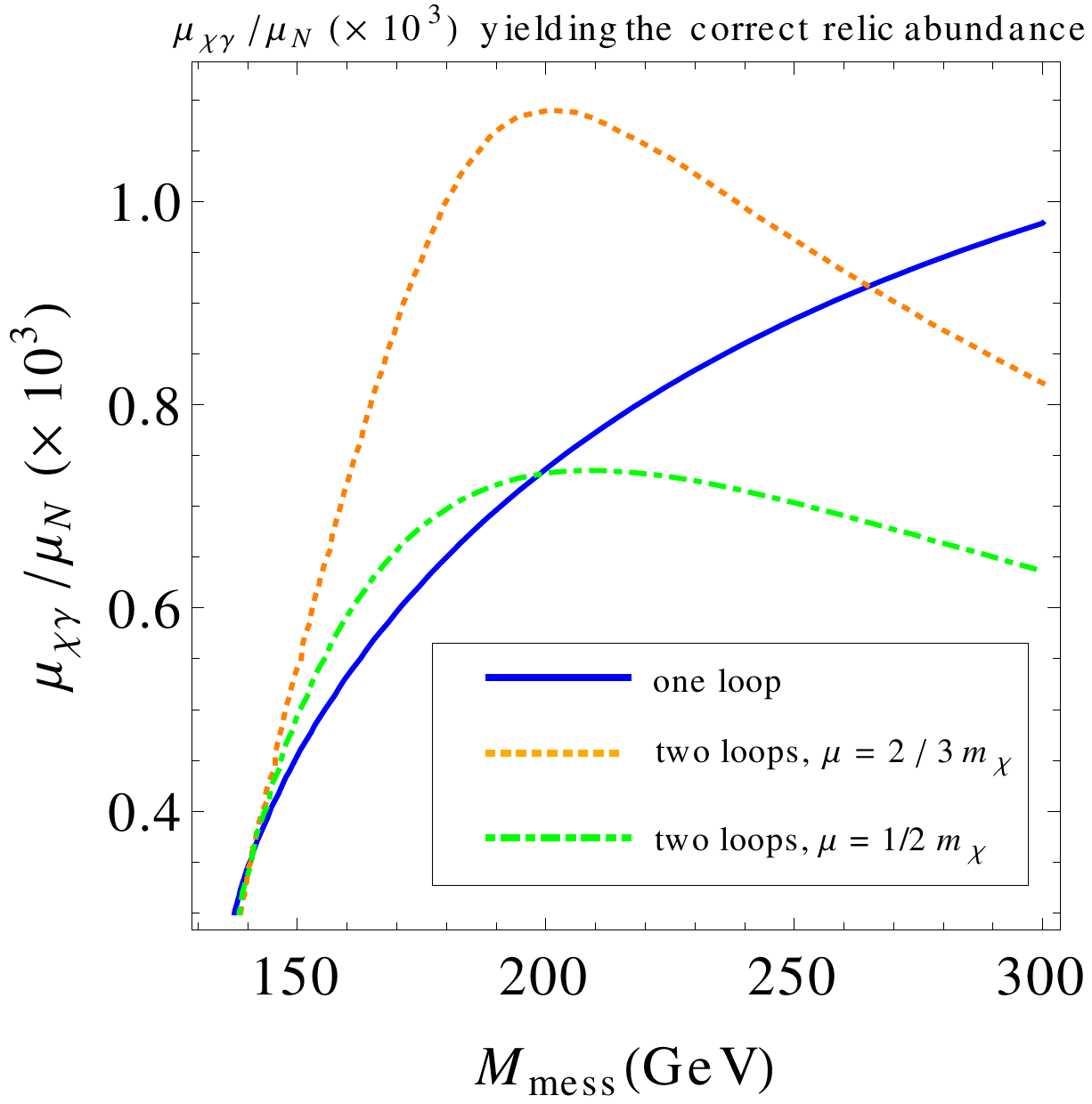}
\includegraphics[width=0.45 \textwidth]{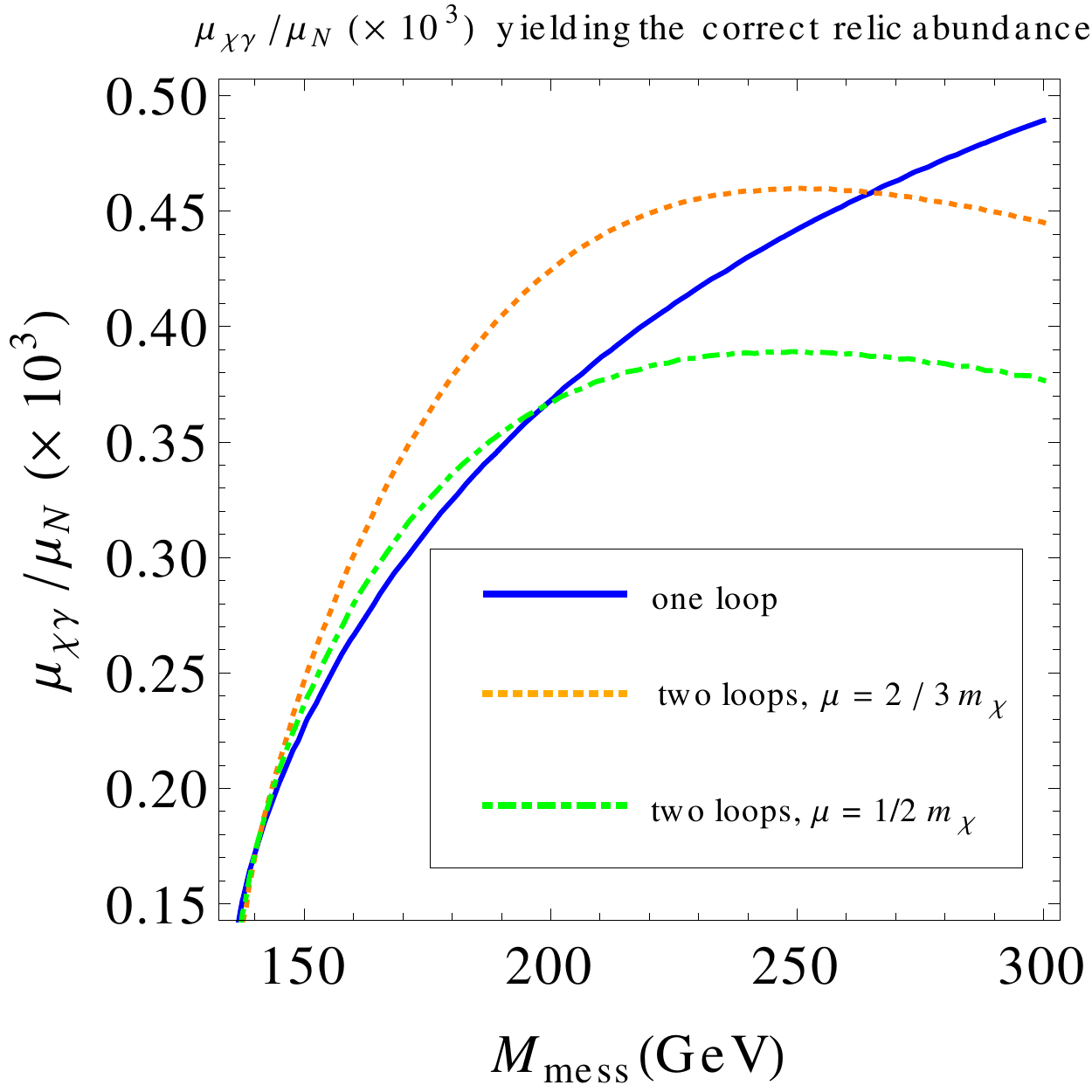}
\end{center}
\caption{The magnetic dipole moment normalized to the nuclear magneton plotted against the messengers' mass fixing the freezeout rate at $\sigma v = 6\times 10^{-26}{\rm cm^3/s}$, for different values of the RG scale $\mu$, and for the case of one family of messengers (top pane) and two families (bottom pane).}
\label{fig:muXg_vs_Mmess}
\end{figure}

At lowest order the direct-detection rate is therefore mostly affected by the magnetic dipole moment contribution. We plot the strength of the magnetic dipole in Fig.~\ref{fig:muXg_vs_Mmess} against the messenger mass, fixing the Yukawa coupling by requiring a freeze-out rate of $\sigma v = 6\times 10^{-26}{\rm cm^3/s}$. For elastic scattering, such a large magnetic moment is strongly excluded by current direct-detection experiments~\cite{Fortin:2011hv,Weiner:2012gm} at the level of $\muX \lesssim 6 \times 10^{-5}\muN$ for $\mX \sim 135\GeV$. However, in the inelastic case with $\delta \sim 100 \keV$ the constraints are considerably weaker and such magnetic moments are within reach of current efforts~\cite{Chang:2010en}. Indeed, one of the original motivations for the MiDM model was to explain the signal seen by the DAMA collaboration~\cite{Bernabei:2010mq}. The magnetic dipole strength needed for that purpose is a little larger $\sim~{\rm few}~\times 10^{-3}\muN$, but it is surprisingly close to the values shown in Fig.~\ref{fig:muXg_vs_Mmess}. In that regard, one should keep in mind that the DAMA experiment suffers from unknown backgrounds that make it difficult to extract the precise overall rate of any putative signal seen in the modulation analysis~\cite{Pradler:2012qt, Pradler:2012bf}.  

The differential cross-sections of Eqs.~(\ref{eqn:sigmaDD}) and (\ref{eqn:sigmaDZ}) are slightly modified in the presence of the form-factors $f_1$ and $F_2$ of Eq.~(\ref{eqn:lowQ_vertex}),
\be
\label{eqn:sDD_2_loop}
\frac{d \sDD}{d \ER}&=& F_2^2\left. \frac{d \sDD}{d \ER}\right|_{f_1=0}, \\\nonumber
\\
\label{eqn:sDZ_2_loop}
\frac{d \sDZ}{d \ER}&=& F_2^2\left. \frac{d \sDZ}{d \ER}\right|_{f_1=0} \\ \nonumber &+& Z_\chi\frac{4\pi \alpha^2 Z^2}{3 v^2}\left( \frac{\muXg}{e} \right)^2\frac{1}{\mX}\left( \frac{\mN}{\mmedf}  F_2  f_1\!+\! \frac{\mX\mN}{6\mmedf^2}  f_1^2 \right).
\ee
It is beyond the scope of the present work to explore in detail the effect of these modifications on all the different direct detection experiments. But inspection of Eqs.~(\ref{eqn:sDD_2_loop}) and~(\ref{eqn:sDZ_2_loop}) and the results presented in the previous section for the form-factors in the low momentum exchange region (see e.g. the one-loop results in Eqs.~(\ref{eqn:f1_1_loop}) and~(\ref{eqn:F2_1_loop}) and the two-loop corrections in Fig.~(\ref{fig:FF_lowQ})), should make it clear that the changes are rather mild unless the messengers are very close in mass to the WIMP. As shown in  Eq.~(\ref{eqn:f1_1_loop}) and Fig.~(\ref{fig:FF_lowQ}) the form-factor $|f_1|$ approaches unity quadratically as the ratio of WIMP mass to messenger mass, $\mX/\mmedf$ diminishes. The dipole form-factor $F_2$ does so only linearly, as in Eq.~(\ref{eqn:f1_1_loop}), but both form-factors appear quadratically in the cross-section and so the overall effect is rather mild. Nevertheless, using the results presented above, it is now 
possible to reliably calculate the corrections to the scattering cross-section in the model defined by the Yukawa coupling Eq.~(\ref{eqn:Yukawa-type}) for any of the different direct detection experiments. 

\subsection{LHC production\label{subsec:LHCprod}}

To close the study of phenomenological aspects of the model, in this section we focus on the cross-section for pair production of the WIMP at the LHC. One should keep in mind that since LHC beam energies are much above the charged messengers mass these states can be produced directly on-shell. Indeed, the messengers are much more easily produced than the WIMP itself, but the LHC phenomenology associated with these states strongly depends on how they decay. This was recently discussed in ref.~\cite{Liu:2013gba} where a systematic study of this possibility was presented. Here we confine ourselves to the production of the WIMP itself. In the limit of degenerate masses for the Majorana WIMPS, the decay products of the heavier state are too soft to be detected, and dark matter may be signaled by unbalanced initial state radiation.  Thus the relevant searches are mono-photon and mono-jet searches, as was discussed in detail in ref.~\cite{Weiner:2012cb}. 

\begin{figure}
\begin{center}
\includegraphics[width=0.45 \textwidth]{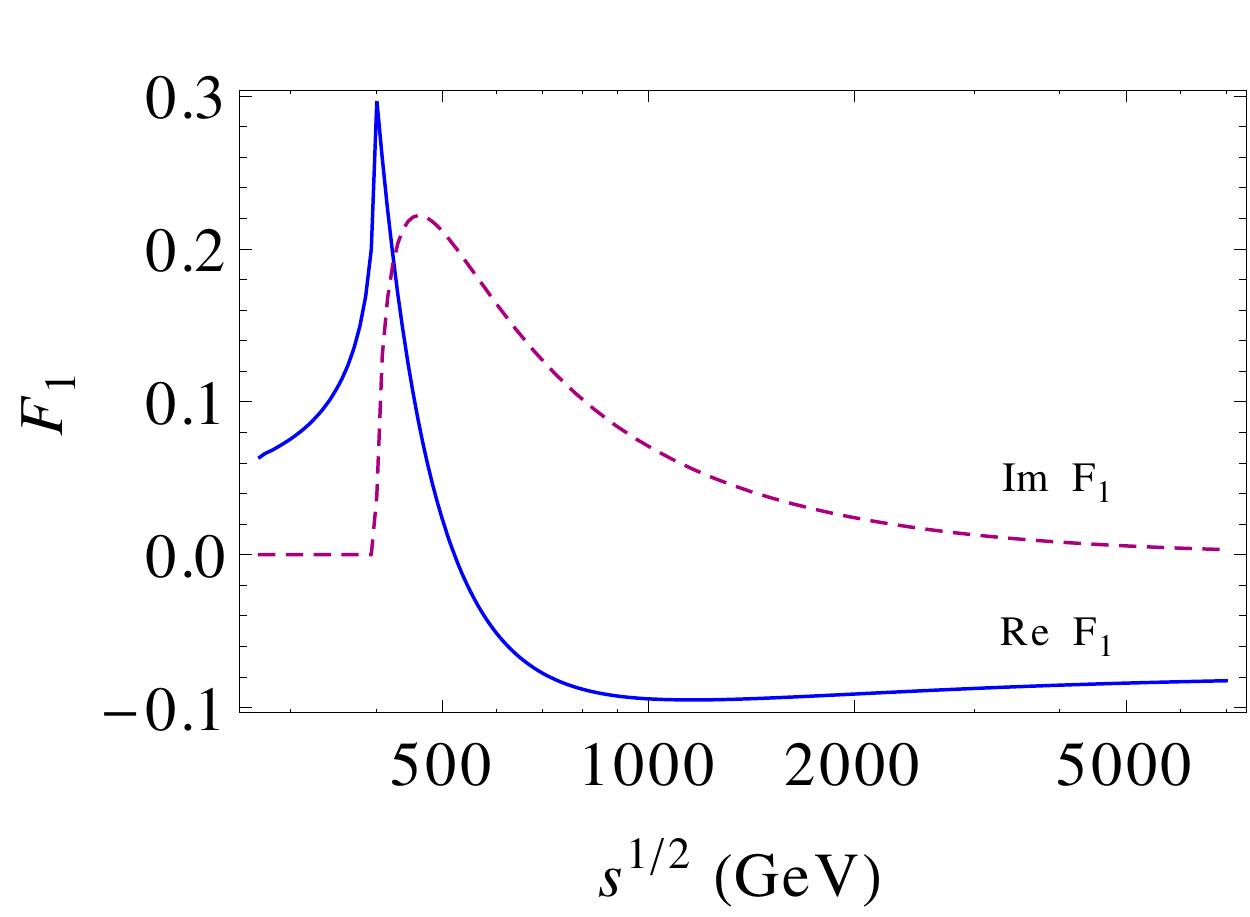}
\includegraphics[width=0.45 \textwidth]{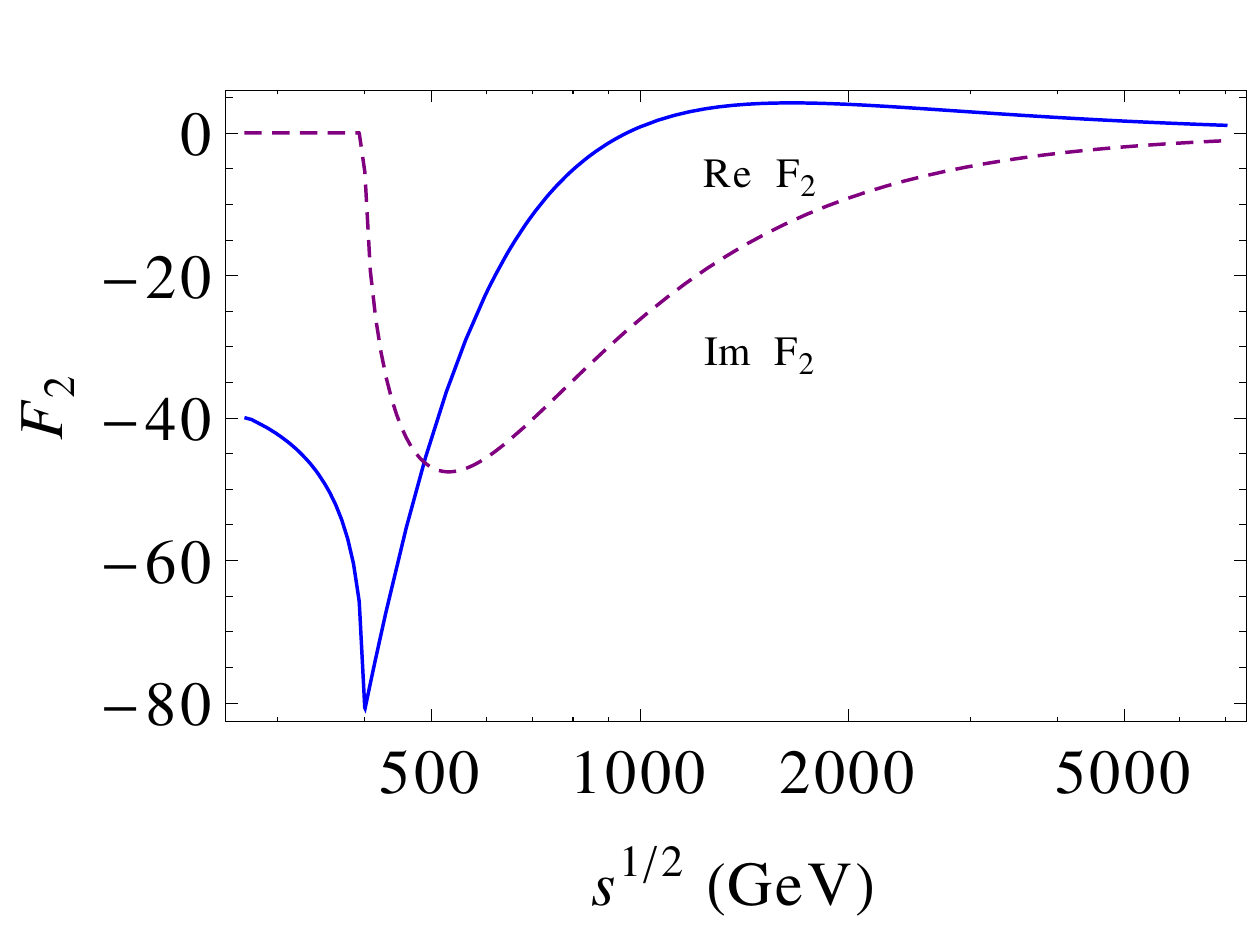}
\end{center}
\caption{Real and imaginary parts of $F_1$ and $F_2$ at one loop as a function of $q^2$, for $\mX=135$ GeV, $\alpha=2$, $\kappa=0$, $M_{mess}=200$ GeV. Notice the threshold singularity at the messenger mass.}
\label{fig:Fs_imaginary}
\end{figure}

 A complete calculation at two-loops is complicated by the fact that the center-of-mass energy $\sqrt{s}$ in the parton frame goes across several singular thresholds of the one and two-loop diagrams, which develop an imaginary part. An accurate computation
can be done by deforming the integration contour in order to avoid the singularities, which is nicely implemented in the software \textsc{SecDec} \cite{Borowka:2012yc}. We were able to do this successfully at one-loop, obtaining the real and imaginary parts of the form factors for the full range of $q^2$ relevant for LHC kinematics; the results are illustrated by Fig.~\ref{fig:Fs_imaginary}, showing the real and imaginary parts of the one-loop form factors as a function of 
$q^2$ for $M_{mess}=200$ GeV and $\alpha=2$; note the singularity at the messenger production threshold at $400$ GeV, beyond which a nonzero imaginary part develops. Due to the daunting computational cost of calculating two-loop diagrams for a wide enough range of values of $\sqrt{s}$ so as to allow for a reliable numerical integration when computing cross-sections, we opted to include two-loop effects only through field renormalization contributions to the scattering amplitudes, as well as in the determination of the coupling yielding the correct relic abundance as in \S~\ref{subsec:relic_abundance}. This approximation is expected to be accurate for the following reasons. As is clear from Fig.~\ref{fig:Fs_imaginary}, the calculation is expected to be dominated by $F_2$ evaluated near the threshold. Now, in section \S~\ref{subsec:timelike} we were already able to compare one and two-loop contributions to $F_2$ near the threshold, when $q^2$ was fixed to $4m_\chi^2$ and the messenger mass was taken to be near $m_\chi$ (bottom pane of Fig.~\ref{fig:F1_timelike}, near the vertical axis). In this situation, the one-loop contribution to $F_2$ peaks while the 2 loop one stays bounded, so that neglecting it should be a safe approximation.

\begin{figure}
\begin{center}
\includegraphics[width=0.45 \textwidth]{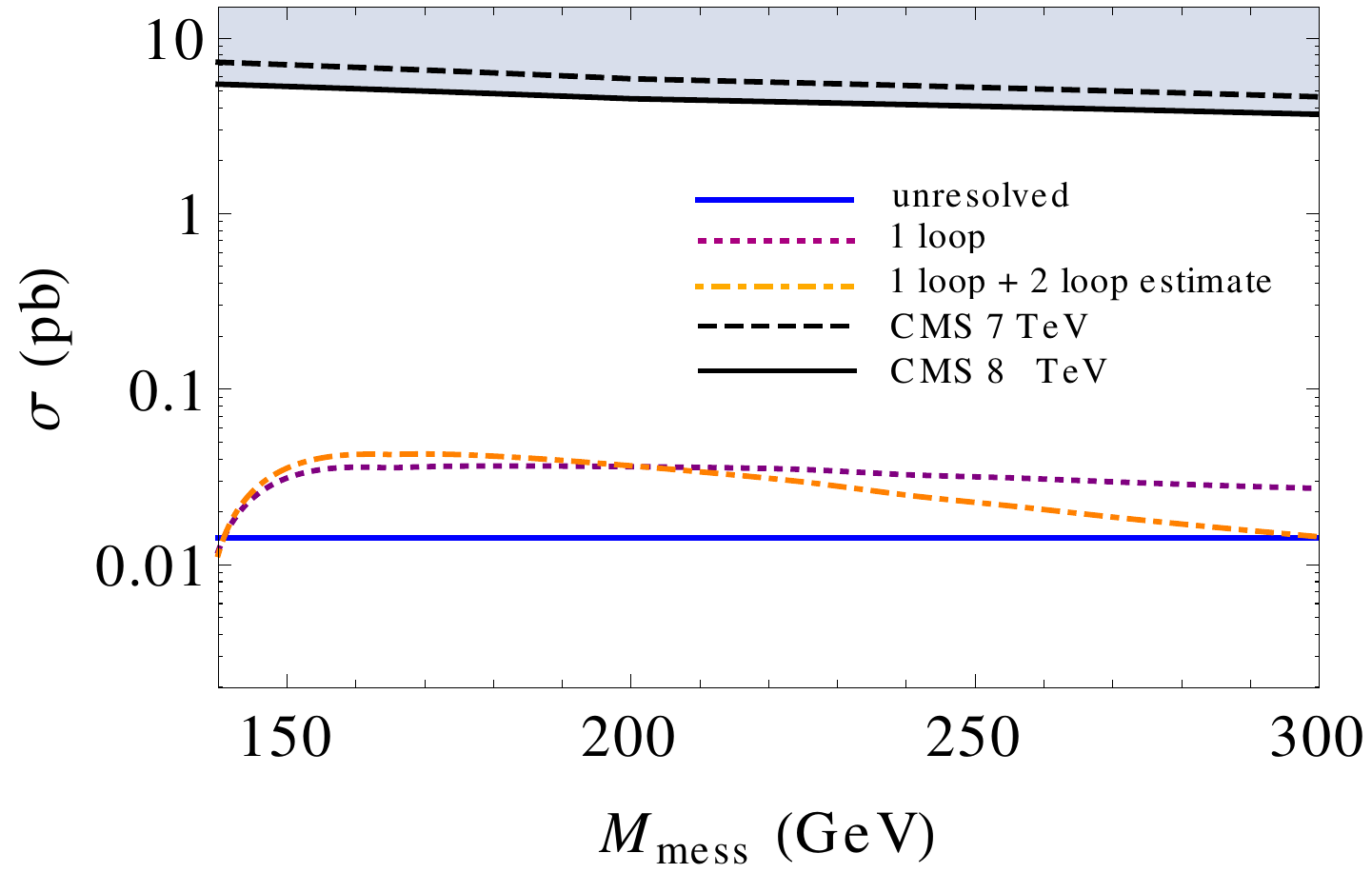}
\end{center}
\caption{LHC production cross section in terms of the messenger mass, in the case of unresolved form factors ($F_1=0,F_2=1$),  at one-loop and also including two-loop effects from the determination of $\alpha_\lambda$ and from field renormalization effects for $\mu=1/2 m_\chi$. The couplings were fixed so as to get the correct relic abundance. Note that the $q$-dependent form factors tend to enhance the cross-section; the dropping of the two/loop curve at large messenger masses is related with two loop corrections demanding lower values of $\alpha_\lambda$ for high messenger masses (see Fig.~\ref{fig:alphaVsMmed}). We also show bounds from CMS monojet searches. }
\label{fig:LHCxsections}
\end{figure}

The LHC cross sections were obtained by integrating over $\sqrt{s}$ the parton  scattering amplitudes in the center-of-mass frame multiplied by the appropriate MSTW parton distribution functions; for this we used the Mathematica package MSTW PDFs \cite{Martin:2009iq}. The resulting cross-sections, with $\lambda$ fixed by 
requiring the correct relic abundance as in \S~\ref{subsec:relic_abundance} (taking $\mu=1/2 m_\chi$ for the two-loop corrections) are plotted in Fig.~\ref{fig:LHCxsections}, where we also include the results with unresolved form factors ($F_1=0, F_2=1$ as in ref.~\cite{Weiner:2012cb}), and also show bounds derived from CMS monojet searches \cite{Chatrchyan:2012me,CMS:rwa}. The latter do not rule out the scenario studied here. It is worth noticing that the $q$-dependent form factors tend to increase the cross-sections with respect to the unresolved case, at least for low messenger masses, which can be attributed to the enhancement of the form factors near the threshold singularity. More details about the calculation of the cross-sections are given in \S~\ref{app:cross_section_formulae}.

%\newpage
\section{Conclusions}
\label{sec:conclusions}

This work was devoted to the investigation of the dominant two-loop contributions to the magnetic dipole moment of a neutral particle that is coupled to charged messengers as defined by the model of Eq.~(\ref{eqn:Lagrangian}). In ref.~\cite{Weiner:2012gm} the authors showed that this model can explain the tentative Fermi line at $135\GeV$ as dark matter annihilating into photons through a loop of the charged messengers. There, it was found that the coupling to the messengers must be sizable, $\alpha_\lambda \sim 1$, in order to obtain the correct phenomenology. Here we showed that this phenomenological success is not spoiled by uncontrolled, $\alpha_\lambda$-dependent corrections at the next order in perturbation theory when  the renormalization scale is chosen appropriately (e.g. $\mu\sim m_\chi/2$ for low messenger masses). A more robust conclusion which is less sensitive to the choice of renormalization scale can only be obtained by including three-loop corrections, a task which is well beyond the scope of the present work. Nevertheless, it is encouraging that with a reasonable choice of renormalization scale the next-to-leading corrections are mild. Including the two-loop corrections, the correct relic abundance can still be obtained when the WIMP annihilates through the dipole moment into SM degrees of freedom. The present day annihilation into di-photons proceeds through the Rayleigh operator~\cite{Weiner:2012gm}, and at least at one-loop order the resulting rate is in agreement with the observed signal. The two-loop computation of the Rayleigh process is left for future work, but based on the current work we expect it to similarly change the one-loop result in only a mild way.  

The phenomenology associated with direct-detection of the WIMP and its direct production at the LHC is also governed by the same one-photon vertex, but at different values of the momentum exchange. We have computed the contribution to these processes as well. Interestingly, the resulting direct-detection rates are within reach of current efforts if the inelastic splitting between the WIMP state and its excited pseudo-Dirac partner is comparable to its kinetic energy in the halo as in the MiDM scenario~\cite{Chang:2010en}.  In the case of production in the LHC the threshold singularities in the form-factors tend to increase the production cross-section of the WIMP and the excited state. If the inelastic splitting is sufficiently large ($\sim 10$'s GeV) then the signal can easily be seen in monophoton searches by making usage of the final-state photon emitted by the excited state as it relaxes to the ground state~\cite{Weiner:2012cb}.  For smaller splittings, this photon is likely too soft to be detected directly, and the signal must be looked for in monojet searches using the initial-state QCD radiation. The limits coming from such searches are currently not sufficiently sensitive to detect the range of dipole strength discussed in this work (see Fig.~\ref{fig:LHCxsections}).

It remains to be seen whether the Fermi line holds up and whether the model considered in this paper has anything to do with dark matter.

\begin{acknowledgments}
IY is supported in part by funds from the Natural Sciences and Engineering Research Council (NSERC) of Canada. Research at the Perimeter Institute is supported in part by the Government of Canada through Industry Canada, and by the Province of Ontario through the Ministry of Research and Information (MRI). CT acknowledges support from the Spanish Government through grant FPA2011-24568.
\end{acknowledgments}

\newpage
\onecolumngrid
\appendix

%%%%%%%%%%%%%%%%
% Appendix A
%%%%%%%%%%%%%%%%

\renewcommand{\theequation}{A-\arabic{equation}}
\setcounter{equation}{0}

\section{}
\label{app:quartic_contribution}

The contribution of the quartic coupling to the magnetic dipole moment appears at two-loops and the relevant diagrams are displayed in Fig.~\ref{fig:all_2_loop_diagrams_Quartic}, with corresponding diagrams where a counter-term is inserted. These diagrams all involve corrections to the scalar propagator since at this order there is no gauge-boson vertex correction. As such, they are particularly easy to compute because the scalar propagator correction does not depend on the momentum of the propagator and so  it factorizes. 
\vspace{1cm}
\be
&~&\label{eqn:Scalar-propagator-1-loop-quartic+ct}
\parbox[t]{5cm}{\vspace{-.5cm}
  \includegraphics[scale=0.6]{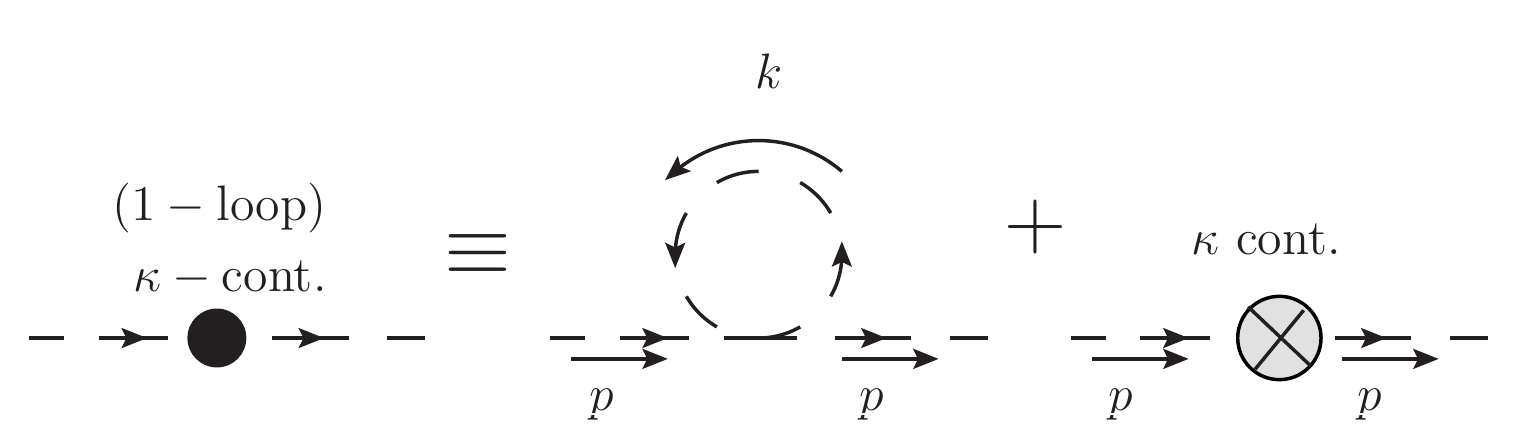} } 
\\\vspace{2.0cm}
 \nonumber  &~& \hspace{3cm}=\frac{-i \kappa \mmeds^2}{16\pi^2}\left(\frac{N_{\meds}+1}{2} \right)\left(\gamma-1 + \log\left(\frac{\mmeds^2}{4\pi \mu^2} \right)\right).
\ee
The two-loop contribution is then,
\vspace{2.0cm}
\be
&~&\label{eqn:2_loop_diagram_9_Quartic}
\parbox[t]{5cm}{\vspace{-1.5cm}
  \includegraphics[scale=0.6]{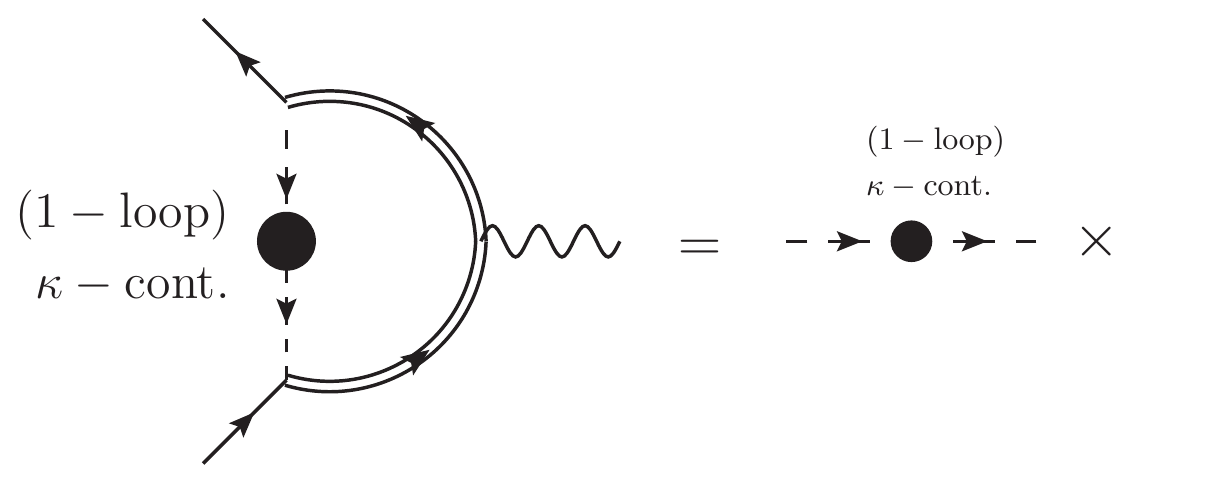} } \hspace{2.0cm}
  \left(\frac{g \lambda^2}{16\pi^2} \right)\int d^3 y \delta(y_1+y_2+y_3-1) \left( \frac{f_9}{\Delta_9} \gamma^\mu+ \frac{g_9}{\Delta_9} \frac{i \sigma^{\mu\nu}q_\nu}{\mmedf} \right),
\ee
where the integral is over the Feynman parameters ${y_1,y_2,y_3}$. Here the one-loop $\kappa$ contribution is given in Eq.~(\ref{eqn:Scalar-propagator-1-loop-quartic+ct}) and
\be
f_9 &=& y_3 \left(\mmedf^2 (1+y_1+y_2)+2 \mmedf \mX y_3+y_3 \left(\mmeds^2+\mX^2 (-1+2 y_3)\right)\right),\\
g_9 &=& y_3 (-1+y_3) (\mmedf+\mX y_3),\\
\Delta_9 &=& \left( (y_1+y_2) \mmedf^2 +y_3 \mmeds^2- (1-y_3) y_3\mX^2 - q^2 y_1 y_2 \right)^2.
\ee
Similarly for diagram (10) in Fig.~\ref{fig:alphaVsMmed} we have
\vspace{2.0cm}
\be
&~&\label{eqn:2_loop_diagram_10_Quartic}
\parbox[t]{5cm}{\vspace{-1.5cm}
  \includegraphics[scale=0.6]{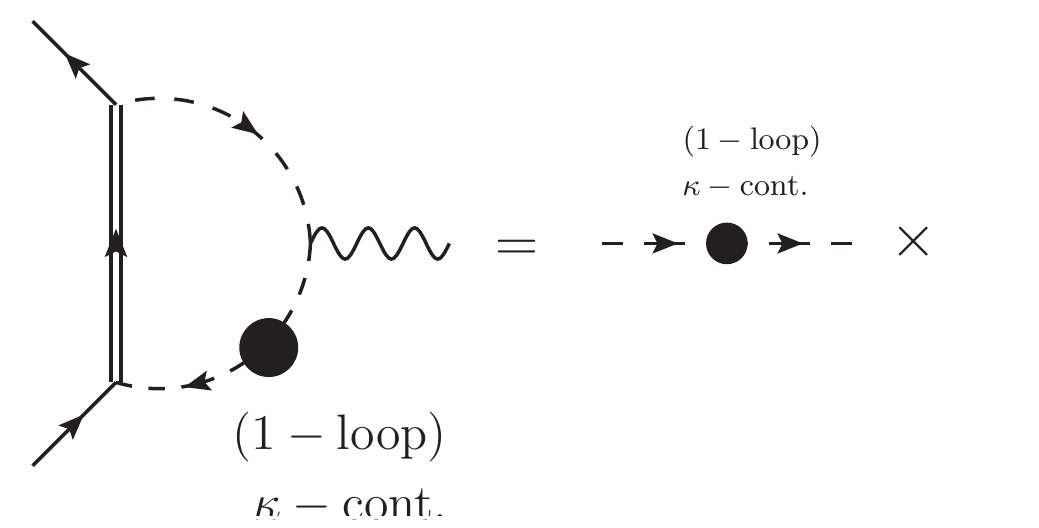} } \hspace{1.0cm}
  \left(\frac{g \lambda^2}{16\pi^2} \right)\int d^3 y \delta(y_1+y_2+y_3-1) \left( \frac{f_{10}}{\Delta_{10}} \gamma^\mu+ \frac{g_{10}}{\Delta_{10}} \frac{i \sigma^{\mu\nu}q_\nu}{\mmedf} \right),
\ee
where
\be
f_{10} &=& -y_2 \left(y_1 \left((\mmedf+\mX)^2-\mX^2 y_1\right)- y_3 y_2q^2+\mmeds^2 (y_2+y_3)\right), \\
g_{10} &=& - y_2 y_1 \Big(\mmedf+\mX(1- y_1)\Big), \\
\Delta_{10} &=& \left( y_1 \mmedf^2  + (y_2+y_3)\mmeds^2  -(1-y_1) y_1  \mX^2  -y_3 y_2 q^2 \right)^2.
\ee
The last diagram in Fig.~\ref{fig:alphaVsMmed}, diagram (11), yields the same contribution as diagram (10). 

%%%%%%%%%%%%%%%%%%%%%%%%%%%%%%%%%%%%%%%%%%%%%%%%%%%%%%%%%%%%%%%%%%%%%%%%%%%%
%%%%%%%%%%%%%%%%%%%%%%%%%%%%%%%%%%%%%%%%%%%%%%%%%%%%%%%%%%%%%%%%%%%%%%%%%%%%
%%%%%%%%%%%%%%%%%%%%%%%%%%%%%%%%%%%%%%%%%%%%%%%%%%%%%%%%%%%%%%%%%%%%%%%%%%%%

\renewcommand{\theequation}{B-\arabic{equation}}
\setcounter{equation}{0}

\section{}
\label{app:cross_section_formulae}
\begin{figure}
\begin{center}
\includegraphics[width=0.45 \textwidth]{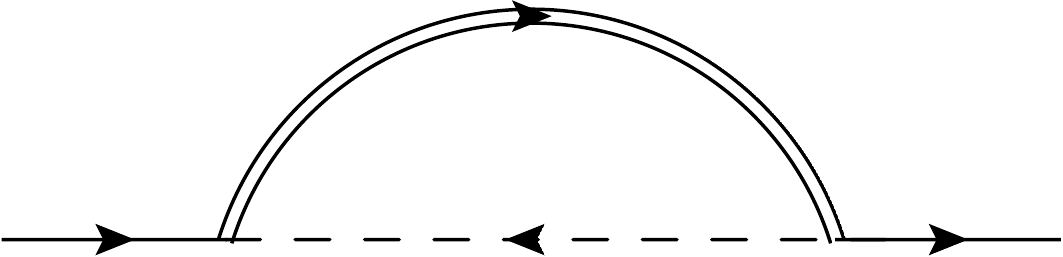}
\end{center}
\caption{Diagram contributing to field and mass renormalization}
\label{fig:chiprop}
\end{figure}

This appendix presents formulae for relevant cross-sections. At two-loops, it is necessary to include one-loop field renormalization factors in the LSZ reduction formula for scattering amplitudes. 
Taking only into account the contributions coming from the coupling $\lambda$, it suffices to consider the field renormalization factor induced on the dark matter field $\chi$ by the finite part of the propagator correction of Fig.~\ref{fig:chiprop}. Evaluation of the diagram for $\mmedf=\mmeds$ yields, after proper renormalization in the minimal subtraction scheme,
\begin{align}\nonumber
 i{\mathcal M}_{FP}(p)=&i\left(\frac{\slashed{p}}{2}+\mmedf\right)\delta Z(p^2,\mu),\\
 %%%%%
 \label{eqn:ZDM}\delta Z(p^2,\mu)=&\frac{\lambda^2 N_m}{8\pi^2}\left(2-
 \gamma_E+\log\frac{4\pi \mu^2}{\mmedf^2}-2\sqrt{\frac{4\mmedf^2-p^2}{p^2}}\text{ArcCsc}\left[2\sqrt{\frac{M^2_f}{p^2}}\right]\right),
\end{align}
where $N_m$ is the number of messenger fields.
This implies that the relation between the mass parameter $\mX$ in the Lagrangian and the physical dark matter mass $m_{\chi,phys}$ is modified to
\begin{align}\label{eqn:masscorrection}
 m^2_{\chi,phys}\left(1+\frac{\delta Z(m^2_{\chi,phys},\mu)}{2}\right)^2=\mX^2\left(1-\frac{\mmedf}{\mX}\delta Z(m^2_{\chi,phys},\mu)\right)^2,
\end{align}
while the field renormalization factor entering the LSZ formula is
\begin{align}\label{eqn:ZDM2}
 Z_\chi(\mu)=\left(1+\frac{\delta Z( m^2_{\chi,phys},\mu)}{2}\right)^{-1}.
\end{align}
In the 2 loop diagrams of Fig.~\ref{fig:all_2_loop_diagrams_Yukawa}, there are propagators of the dark matter field $\chi$ whose masses should be corrected according to Eqs.~\eqref{eqn:ZDM} and \eqref{eqn:masscorrection}, while the mass terms coming from phase-space integration or on-shell relations should be kept at the physical value, which was fixed at 135 GeV. To avoid this complication in the evaluation of the integrals, we used a single value for the mass and substituted the DM propagator by the pole of the one-loop contribution,
\begin{align*}
 \frac{\slashed{p}+\mX}{p^2-\mX^2}\rightarrow  Z_\chi(m^2_{\chi,phys
 },\mu)\frac{\slashed{p}+m_{\chi,phys
 }}{p^2-m_{\chi,phys}^2}.
\end{align*}
This is equivalent to including some 3 loop corrections.

The non-relativistic annihilation cross-section of dark matter particles $\chi$ and $\chi^*$  into light fermions by s-schannel exchange of $Z$ and $\gamma$ is given, in a nonrelativistic limit in which $\chi$ and $\chi^*$ are taken as degenerate with mass $m$, by
\begin{align*} 
&\sigma(\chi\chi^*\rightarrow f f)v=
\frac{Z_\chi}{512\pi  m^6 \left(m_Z^2-4 m^2\right)^2}\Big(e^2 q_f^2 \sec ^2\theta_w \left(16 m^4 \left(a_f^2+v_f^2+1\right)+\cos (2 \theta_W) \left(-16 m^4 \left(a_f^2+v_f^2-1\right)\right.\right.\\
%%%%
&\left.\left.-8 m^2 m_Z^2+m_Z^4\right)+8 m^2 v_f \left(4 m^2-m_Z^2\right) \sin (2 \theta_W)-8 m^2 m_Z^2+m_Z^4\right) \left(16 e^2 F_1^2 m^4+8 F_2 m^2 \mu_\chi \cos \theta_w \left(8 F_2 m^4 \mu_\chi \cos \theta_w\right.\right.\\
%%%%
&\left.\left.-8 e F_1 m^3\right)\right)\Big),
\end{align*}
where $\theta_W$ is Weinberg's angle, $q_f$ is the charge of the final fermion state, and $a_f$ and $v_f$ denote the fermion field's axial and vector couplings to the $Z$ boson normalized to the  coupling to the photon. $F_1$ and $F_2$ designate the form factors evaluated at $q^2=4 m^2$.

Similarly, the nonrelativistic annihilations into vector bosons $\gamma,Z$ are given by
\begin{align*}
 \sigma(\chi\chi^*\rightarrow \gamma\gamma)v=&Z_\chi\frac{F_2(0)^4 \,{m}^2 {\mu_\chi}^4  \cos ^4\theta_w}{4\pi },\\
 %%%%%%%
%%%%%%%
  \sigma(\chi\chi^*\rightarrow \gamma Z)v=&Z_\chi
 \frac{F_2(0)^2 {\mu_\chi}^2 \left(4 {m}^2-m_Z^2\right)\sin ^2(\theta_w) \left(F_2(m^2_Z) {\mu_\chi} \cos (\theta_w) \left(4 m^2+m_Z^2\right)-4 e F_1(m^2_Z) {m}\right)^2}{128 \pi  {m}^4 }.
\end{align*}

In the above formulae, we used the fact that $F_1(0)=0$ (there is no long-range electromagnetic coupling for dark matter), and the form factors were assumed to be real, which is a valid assumption for nonrelativistic values of $q^2=4 m^2<4 \mmedf^2$ for which no threshold singularities are reached.
More care has to be taken in the case of LHC production cross sections, which involve an integration over the partons' center of mass energy, so that thresholds for the production of the messenger fields $\psi$ and $\phi$ are crossed, causing the form factors to get an imaginary part. The differential production cross-section of $\chi,\chi^*$ by massless quarks is, this time admitting nondegenarate masses $m$ and $m_*$,
\begin{align*}
& \frac{d\sigma(f f\rightarrow \chi\chi^*)}{dt}=Z_\chi\frac{1}{8 \pi  s^4}e^2 q_f^2\Big(e^2 |F_1|^2 (s^2 + m^2 (2 m_*^2 - s - 2 t) + m_*^2 (s - 2 t) + 2 s t + 
   2 t^2)\\
   %%%%%
   &+ \frac{1}{(m_Z^2 - 
          s)^2} s (2 e\, \text{Re}(F_1 F_2^*) s ((m m_*^2 s (a_f^2 + v_f^2) - 
                    m_*^3 s (a_f^2 + v_f^2) - 
         m (m^2 - s) s (a_f^2 + v_f^2) + 
                    m_* (-2 m_Z^4 + 4 m_Z^2 s + \\
   %%%
   &s (-2 s + (m^2 + s) (a_f^2 + v_f^2)))) \cos(
                  \theta_w) + (m + m_*) ((m - m_*)^2 - 
                    s) (s (a_f^2 + v_f^2) \sec(\theta_w) + 
                    2 (-m_Z^2 + s) v_f \sin(\theta_w))) \mu_\chi \\
   %%%
   &+     |F_2|^2 (-(m_Z^2 - s)^2 (m^4 + m_*^4 + 2 t (s + t) - 
         m^2 (s + 2 t) -
                     m_*^2 (3 s + 2 t)) \cos(\theta_w)^2 + 
      s (-m^4 - m_*^4 + 2 m m_* s - 2 t (s + t) \\
   %%%%
   &+ m^2 (s + 2 t) + 
                    m_*^2 (s + 2 t)) \sin(       
        \theta_w) (2 (-m_Z^2 + s) v_f \cos(\theta_w) + s (a_f^2 + v_f^2) \sin(\theta_w))) 
\mu_\chi^2 + 
        e^2 |F_1|^2 (2 m m_* s + s^2 + m^2 (2 m_*^2\\
  %%%%%
  &      - s - 2 t) + 
      2 s t + 
              2 t^2 - m_*^2 (s + 2 t)) \tan(
            \theta_w) (2 (-m_Z^2 + s) v_f + s (a_f^2 + v_f^2) \tan(\theta_w)))\Big),
\end{align*}
where the Mandelstam variable $t$ is $t=(p_a-p_\chi)^2$, $p_a$ being the momentum of the incoming fermion and $p_\chi$ that of the particle with mass $m$.

%%%%%%%%%%%%%%%%%%%%%%%%%%%%%%%%%%%%%%%%%%%%%%%%%%%%%%%%%%%%%%%%%%%%%%%%%%%%
%%%%%%%%%%%%%%%%%%%%%%%%%%%%%%%%%%%%%%%%%%%%%%%%%%%%%%%%%%%%%%%%%%%%%%%%%%%%
%%%%%%%%%%%%%%%%%%%%%%%%%%%%%%%%%%%%%%%%%%%%%%%%%%%%%%%%%%%%%%%%%%%%%%%%%%%%

\renewcommand{\theequation}{C-\arabic{equation}}
\setcounter{equation}{0}

\section{}
\label{app:one-loop}
This appendix reproduces the formulae for the one-loop form factors that were obtained in ref.~\cite{Weiner:2012cb}. The form-factors are given by the following integrals:
\begin{align*}
 F_1(q^2;\mX,\mmedf,\mmeds)=&-\frac{g'\lambda^2}{16\pi^2}\,\int dx~dy~dz\left(1+\frac{2\mX(\mmedf+(1-z)\mX)}{\Delta_s}-\frac{(\mmedf+z\mX)^2+xy\,q^2}{\Delta_f}+\log\left(\frac{\Delta_f}{\Delta_s}\right)\right),\\
 %%%%%%%%%%%
 F_2(q^2;\mX,\mmedf,\mmeds)=&-2\mmedf\,\int dx~dy~dz\left(1+\frac{z(\mmedf+(1-z)\mX)}{\Delta_s}+\frac{2x(\mmedf+z\mX)}{\Delta_f}\right),
 \end{align*}
where the integrals are done under the constraint that $x+y+z = 1$. The denominators are given by
\be
\Delta_f &=& z \mmeds^2 + (x+y) \mmedf^2 -z(1-z)\mX^2 -x y ~q^2, \\
\Delta_s &=& z \mmedf^2 + (x+y) \mmeds^2 -z(1-z)\mX^2 -x y ~q^2.
\ee
In the limit of heavy messenger mass these form-factors are 
\be
F_1(q^2) &=& \frac{\mu_\chi q^2}{6\mmedf}\left(\frac{2r^2 \left(3 r^2-3-\left(2+r^2\right) \log r^2\right)}{ \left(1-r^2\right)^2}\right), \\
F_2(q^2) &=& -\frac{2 r^2 \left(r^2-1-\log r^2\right)}{\left(1-r^2\right)^2},
\label{eqn:F2nrlimit}
\ee
where $r = \mmedf/\mmeds$. 

%%%%%%%%%%%%%%%%%%%%%%%%%%%%%%%%%%%%%%%%%%%%%%%%%%%%%%%%%%%%%%%%%%%%%%%%%%%%%%%%%%%%%%%%%%%%%%%%%%%%
%%%%%%%%%%%%%%%%%%%%%%%%%%%%%%%%%%%%%%%%%%%%%%%%%%%%%%%%%%%%%%%%%%%%%%%%%%%%%%%%%%%%%%%%%%%%%%%%%%%%
\bibliography{bib-MDM_2-loop}

\begin{thebibliography}{44}%
\makeatletter
\providecommand \@ifxundefined [1]{%
 \@ifx{#1\undefined}
}%
\providecommand \@ifnum [1]{%
 \ifnum #1\expandafter \@firstoftwo
 \else \expandafter \@secondoftwo
 \fi
}%
\providecommand \@ifx [1]{%
 \ifx #1\expandafter \@firstoftwo
 \else \expandafter \@secondoftwo
 \fi
}%
\providecommand \natexlab [1]{#1}%
\providecommand \enquote  [1]{``#1''}%
\providecommand \bibnamefont  [1]{#1}%
\providecommand \bibfnamefont [1]{#1}%
\providecommand \citenamefont [1]{#1}%
\providecommand \href@noop [0]{\@secondoftwo}%
\providecommand \href [0]{\begingroup \@sanitize@url \@href}%
\providecommand \@href[1]{\@@startlink{#1}\@@href}%
\providecommand \@@href[1]{\endgroup#1\@@endlink}%
\providecommand \@sanitize@url [0]{\catcode `\\12\catcode `\$12\catcode
  `\&12\catcode `\#12\catcode `\^12\catcode `\_12\catcode `\%12\relax}%
\providecommand \@@startlink[1]{}%
\providecommand \@@endlink[0]{}%
\providecommand \url  [0]{\begingroup\@sanitize@url \@url }%
\providecommand \@url [1]{\endgroup\@href {#1}{\urlprefix }}%
\providecommand \urlprefix  [0]{URL }%
\providecommand \Eprint [0]{\href }%
\providecommand \doibase [0]{http://dx.doi.org/}%
\providecommand \selectlanguage [0]{\@gobble}%
\providecommand \bibinfo  [0]{\@secondoftwo}%
\providecommand \bibfield  [0]{\@secondoftwo}%
\providecommand \translation [1]{[#1]}%
\providecommand \BibitemOpen [0]{}%
\providecommand \bibitemStop [0]{}%
\providecommand \bibitemNoStop [0]{.\EOS\space}%
\providecommand \EOS [0]{\spacefactor3000\relax}%
\providecommand \BibitemShut  [1]{\csname bibitem#1\endcsname}%
\let\auto@bib@innerbib\@empty
%</preamble>
\bibitem [{\citenamefont {Kusenko}(2009)}]{Kusenko:2009up}%
  \BibitemOpen
  \bibfield  {author} {\bibinfo {author} {\bibfnamefont {A.}~\bibnamefont
  {Kusenko}},\ }\href {\doibase 10.1016/j.physrep.2009.07.004} {\bibfield
  {journal} {\bibinfo  {journal} {Phys.Rept.}\ }\textbf {\bibinfo {volume}
  {481}},\ \bibinfo {pages} {1} (\bibinfo {year} {2009})},\ \Eprint
  {http://arxiv.org/abs/0906.2968} {arXiv:0906.2968 [hep-ph]} \BibitemShut
  {NoStop}%
%%CITATION = ARXIV:0906.2968;%%
\bibitem [{\citenamefont {Bagnasco}\ \emph {et~al.}(1994)\citenamefont
  {Bagnasco}, \citenamefont {Dine},\ and\ \citenamefont
  {Thomas}}]{Bagnasco:1993st}%
  \BibitemOpen
  \bibfield  {author} {\bibinfo {author} {\bibfnamefont {J.}~\bibnamefont
  {Bagnasco}}, \bibinfo {author} {\bibfnamefont {M.}~\bibnamefont {Dine}}, \
  and\ \bibinfo {author} {\bibfnamefont {S.~D.}\ \bibnamefont {Thomas}},\
  }\href {\doibase 10.1016/0370-2693(94)90830-3} {\bibfield  {journal}
  {\bibinfo  {journal} {Phys. Lett.}\ }\textbf {\bibinfo {volume} {B320}},\
  \bibinfo {pages} {99} (\bibinfo {year} {1994})},\ \Eprint
  {http://arxiv.org/abs/hep-ph/9310290} {arXiv:hep-ph/9310290} \BibitemShut
  {NoStop}%
%%CITATION = HEP-PH/9310290;%%
\bibitem [{\citenamefont {Pospelov}\ and\ \citenamefont {ter
  Veldhuis}(2000)}]{Pospelov:2000bq}%
  \BibitemOpen
  \bibfield  {author} {\bibinfo {author} {\bibfnamefont {M.}~\bibnamefont
  {Pospelov}}\ and\ \bibinfo {author} {\bibfnamefont {T.}~\bibnamefont {ter
  Veldhuis}},\ }\href {\doibase 10.1016/S0370-2693(00)00358-0} {\bibfield
  {journal} {\bibinfo  {journal} {Phys. Lett.}\ }\textbf {\bibinfo {volume}
  {B480}},\ \bibinfo {pages} {181} (\bibinfo {year} {2000})},\ \Eprint
  {http://arxiv.org/abs/hep-ph/0003010} {arXiv:hep-ph/0003010} \BibitemShut
  {NoStop}%
%%CITATION = HEP-PH/0003010;%%
\bibitem [{\citenamefont {Sigurdson}\ \emph {et~al.}(2004)\citenamefont
  {Sigurdson}, \citenamefont {Doran}, \citenamefont {Kurylov}, \citenamefont
  {Caldwell},\ and\ \citenamefont {Kamionkowski}}]{Sigurdson:2004zp}%
  \BibitemOpen
  \bibfield  {author} {\bibinfo {author} {\bibfnamefont {K.}~\bibnamefont
  {Sigurdson}}, \bibinfo {author} {\bibfnamefont {M.}~\bibnamefont {Doran}},
  \bibinfo {author} {\bibfnamefont {A.}~\bibnamefont {Kurylov}}, \bibinfo
  {author} {\bibfnamefont {R.~R.}\ \bibnamefont {Caldwell}}, \ and\ \bibinfo
  {author} {\bibfnamefont {M.}~\bibnamefont {Kamionkowski}},\ }\href {\doibase
  10.1103/PhysRevD.70.083501} {\bibfield  {journal} {\bibinfo  {journal} {Phys.
  Rev.}\ }\textbf {\bibinfo {volume} {D70}},\ \bibinfo {pages} {083501}
  (\bibinfo {year} {2004})},\ \Eprint {http://arxiv.org/abs/astro-ph/0406355}
  {arXiv:astro-ph/0406355} \BibitemShut {NoStop}%
%%CITATION = ASTRO-PH/0406355;%%
\bibitem [{\citenamefont {Gardner}(2009)}]{Gardner:2008yn}%
  \BibitemOpen
  \bibfield  {author} {\bibinfo {author} {\bibfnamefont {S.}~\bibnamefont
  {Gardner}},\ }\href {\doibase 10.1103/PhysRevD.79.055007} {\bibfield
  {journal} {\bibinfo  {journal} {Phys. Rev.}\ }\textbf {\bibinfo {volume}
  {D79}},\ \bibinfo {pages} {055007} (\bibinfo {year} {2009})},\ \Eprint
  {http://arxiv.org/abs/0811.0967} {arXiv:0811.0967 [hep-ph]} \BibitemShut
  {NoStop}%
%%CITATION = 0811.0967;%%
\bibitem [{\citenamefont {Masso}\ \emph {et~al.}(2009)\citenamefont {Masso},
  \citenamefont {Mohanty},\ and\ \citenamefont {Rao}}]{Masso:2009mu}%
  \BibitemOpen
  \bibfield  {author} {\bibinfo {author} {\bibfnamefont {E.}~\bibnamefont
  {Masso}}, \bibinfo {author} {\bibfnamefont {S.}~\bibnamefont {Mohanty}}, \
  and\ \bibinfo {author} {\bibfnamefont {S.}~\bibnamefont {Rao}},\ }\href
  {\doibase 10.1103/PhysRevD.80.036009} {\bibfield  {journal} {\bibinfo
  {journal} {Phys. Rev.}\ }\textbf {\bibinfo {volume} {D80}},\ \bibinfo {pages}
  {036009} (\bibinfo {year} {2009})},\ \Eprint {http://arxiv.org/abs/0906.1979}
  {arXiv:0906.1979 [hep-ph]} \BibitemShut {NoStop}%
%%CITATION = 0906.1979;%%
\bibitem [{\citenamefont {Cho}\ \emph {et~al.}(2010)\citenamefont {Cho},
  \citenamefont {Huh}, \citenamefont {Kim}, \citenamefont {Kim},\ and\
  \citenamefont {Kyae}}]{Cho:2010br}%
  \BibitemOpen
  \bibfield  {author} {\bibinfo {author} {\bibfnamefont {W.~S.}\ \bibnamefont
  {Cho}}, \bibinfo {author} {\bibfnamefont {J.-H.}\ \bibnamefont {Huh}},
  \bibinfo {author} {\bibfnamefont {I.-W.}\ \bibnamefont {Kim}}, \bibinfo
  {author} {\bibfnamefont {J.~E.}\ \bibnamefont {Kim}}, \ and\ \bibinfo
  {author} {\bibfnamefont {B.}~\bibnamefont {Kyae}},\ }\href {\doibase
  10.1016/j.physletb.2010.02.081} {\bibfield  {journal} {\bibinfo  {journal}
  {Phys. Lett.}\ }\textbf {\bibinfo {volume} {B687}},\ \bibinfo {pages} {6}
  (\bibinfo {year} {2010})},\ \Eprint {http://arxiv.org/abs/1001.0579}
  {arXiv:1001.0579 [hep-ph]} \BibitemShut {NoStop}%
%%CITATION = 1001.0579;%%
\bibitem [{\citenamefont {An}\ \emph {et~al.}(2010)\citenamefont {An},
  \citenamefont {Chen}, \citenamefont {Mohapatra}, \citenamefont {Nussinov},\
  and\ \citenamefont {Zhang}}]{An:2010kc}%
  \BibitemOpen
  \bibfield  {author} {\bibinfo {author} {\bibfnamefont {H.}~\bibnamefont
  {An}}, \bibinfo {author} {\bibfnamefont {S.-L.}\ \bibnamefont {Chen}},
  \bibinfo {author} {\bibfnamefont {R.~N.}\ \bibnamefont {Mohapatra}}, \bibinfo
  {author} {\bibfnamefont {S.}~\bibnamefont {Nussinov}}, \ and\ \bibinfo
  {author} {\bibfnamefont {Y.}~\bibnamefont {Zhang}},\ }\href@noop {} {\
  (\bibinfo {year} {2010})},\ \Eprint {http://arxiv.org/abs/1004.3296}
  {arXiv:1004.3296 [hep-ph]} \BibitemShut {NoStop}%
%%CITATION = 1004.3296;%%
\bibitem [{\citenamefont {McDermott}\ \emph {et~al.}(2011)\citenamefont
  {McDermott}, \citenamefont {Yu},\ and\ \citenamefont
  {Zurek}}]{McDermott:2010pa}%
  \BibitemOpen
  \bibfield  {author} {\bibinfo {author} {\bibfnamefont {S.~D.}\ \bibnamefont
  {McDermott}}, \bibinfo {author} {\bibfnamefont {H.-B.}\ \bibnamefont {Yu}}, \
  and\ \bibinfo {author} {\bibfnamefont {K.~M.}\ \bibnamefont {Zurek}},\ }\href
  {\doibase 10.1103/PhysRevD.83.063509} {\bibfield  {journal} {\bibinfo
  {journal} {Phys.Rev.}\ }\textbf {\bibinfo {volume} {D83}},\ \bibinfo {pages}
  {063509} (\bibinfo {year} {2011})},\ \Eprint {http://arxiv.org/abs/1011.2907}
  {arXiv:1011.2907 [hep-ph]} \BibitemShut {NoStop}%
%%CITATION = ARXIV:1011.2907;%%
\bibitem [{\citenamefont {Chang}\ \emph {et~al.}(2010)\citenamefont {Chang},
  \citenamefont {Weiner},\ and\ \citenamefont {Yavin}}]{Chang:2010en}%
  \BibitemOpen
  \bibfield  {author} {\bibinfo {author} {\bibfnamefont {S.}~\bibnamefont
  {Chang}}, \bibinfo {author} {\bibfnamefont {N.}~\bibnamefont {Weiner}}, \
  and\ \bibinfo {author} {\bibfnamefont {I.}~\bibnamefont {Yavin}},\ }\href
  {\doibase 10.1103/PhysRevD.82.125011} {\bibfield  {journal} {\bibinfo
  {journal} {Phys.Rev.}\ }\textbf {\bibinfo {volume} {D82}},\ \bibinfo {pages}
  {125011} (\bibinfo {year} {2010})},\ \Eprint {http://arxiv.org/abs/1007.4200}
  {arXiv:1007.4200 [hep-ph]} \BibitemShut {NoStop}%
%%CITATION = ARXIV:1007.4200;%%
\bibitem [{\citenamefont {Banks}\ \emph {et~al.}(2010)\citenamefont {Banks},
  \citenamefont {Fortin},\ and\ \citenamefont {Thomas}}]{Banks:2010eh}%
  \BibitemOpen
  \bibfield  {author} {\bibinfo {author} {\bibfnamefont {T.}~\bibnamefont
  {Banks}}, \bibinfo {author} {\bibfnamefont {J.-F.}\ \bibnamefont {Fortin}}, \
  and\ \bibinfo {author} {\bibfnamefont {S.}~\bibnamefont {Thomas}},\
  }\href@noop {} {\  (\bibinfo {year} {2010})},\ \Eprint
  {http://arxiv.org/abs/1007.5515} {arXiv:1007.5515 [hep-ph]} \BibitemShut
  {NoStop}%
%%CITATION = ARXIV:1007.5515;%%
\bibitem [{\citenamefont {Goodman}\ \emph {et~al.}(2011)\citenamefont
  {Goodman}, \citenamefont {Ibe}, \citenamefont {Rajaraman}, \citenamefont
  {Shepherd}, \citenamefont {Tait} \emph {et~al.}}]{Goodman:2010qn}%
  \BibitemOpen
  \bibfield  {author} {\bibinfo {author} {\bibfnamefont {J.}~\bibnamefont
  {Goodman}}, \bibinfo {author} {\bibfnamefont {M.}~\bibnamefont {Ibe}},
  \bibinfo {author} {\bibfnamefont {A.}~\bibnamefont {Rajaraman}}, \bibinfo
  {author} {\bibfnamefont {W.}~\bibnamefont {Shepherd}}, \bibinfo {author}
  {\bibfnamefont {T.~M.}\ \bibnamefont {Tait}},  \emph {et~al.},\ }\href
  {\doibase 10.1016/j.nuclphysb.2010.10.022} {\bibfield  {journal} {\bibinfo
  {journal} {Nucl.Phys.}\ }\textbf {\bibinfo {volume} {B844}},\ \bibinfo
  {pages} {55} (\bibinfo {year} {2011})},\ \Eprint
  {http://arxiv.org/abs/1009.0008} {arXiv:1009.0008 [hep-ph]} \BibitemShut
  {NoStop}%
%%CITATION = ARXIV:1009.0008;%%
\bibitem [{\citenamefont {Fortin}\ and\ \citenamefont
  {Tait}(2012)}]{Fortin:2011hv}%
  \BibitemOpen
  \bibfield  {author} {\bibinfo {author} {\bibfnamefont {J.-F.}\ \bibnamefont
  {Fortin}}\ and\ \bibinfo {author} {\bibfnamefont {T.~M.}\ \bibnamefont
  {Tait}},\ }\href {\doibase 10.1103/PhysRevD.85.063506} {\bibfield  {journal}
  {\bibinfo  {journal} {Phys.Rev.}\ }\textbf {\bibinfo {volume} {D85}},\
  \bibinfo {pages} {063506} (\bibinfo {year} {2012})},\ \bibinfo {note} {11
  pages, 2 figures, extended discussion, added references, conclusion
  unchanged},\ \Eprint {http://arxiv.org/abs/1103.3289} {arXiv:1103.3289
  [hep-ph]} \BibitemShut {NoStop}%
%%CITATION = ARXIV:1103.3289;%%
\bibitem [{\citenamefont {Del~Nobile}\ \emph {et~al.}(2012)\citenamefont
  {Del~Nobile}, \citenamefont {Kouvaris}, \citenamefont {Panci}, \citenamefont
  {Sannino},\ and\ \citenamefont {Virkajarvi}}]{DelNobile:2012tx}%
  \BibitemOpen
  \bibfield  {author} {\bibinfo {author} {\bibfnamefont {E.}~\bibnamefont
  {Del~Nobile}}, \bibinfo {author} {\bibfnamefont {C.}~\bibnamefont
  {Kouvaris}}, \bibinfo {author} {\bibfnamefont {P.}~\bibnamefont {Panci}},
  \bibinfo {author} {\bibfnamefont {F.}~\bibnamefont {Sannino}}, \ and\
  \bibinfo {author} {\bibfnamefont {J.}~\bibnamefont {Virkajarvi}},\
  }\href@noop {} {\  (\bibinfo {year} {2012})},\ \Eprint
  {http://arxiv.org/abs/1203.6652} {arXiv:1203.6652 [hep-ph]} \BibitemShut
  {NoStop}%
%%CITATION = ARXIV:1203.6652;%%
\bibitem [{\citenamefont {Ackermann}\ \emph {et~al.}(2012)\citenamefont
  {Ackermann} \emph {et~al.}}]{Ackermann:2012qk}%
  \BibitemOpen
  \bibfield  {author} {\bibinfo {author} {\bibfnamefont {M.}~\bibnamefont
  {Ackermann}} \emph {et~al.} (\bibinfo {collaboration} {LAT Collaboration}),\
  }\href {\doibase 10.1103/PhysRevD.86.022002} {\bibfield  {journal} {\bibinfo
  {journal} {Phys.Rev.}\ }\textbf {\bibinfo {volume} {D86}},\ \bibinfo {pages}
  {022002} (\bibinfo {year} {2012})},\ \Eprint {http://arxiv.org/abs/1205.2739}
  {arXiv:1205.2739 [astro-ph.HE]} \BibitemShut {NoStop}%
%%CITATION = ARXIV:1205.2739;%%
\bibitem [{\citenamefont {Bringmann}\ \emph {et~al.}(2012)\citenamefont
  {Bringmann}, \citenamefont {Huang}, \citenamefont {Ibarra}, \citenamefont
  {Vogl},\ and\ \citenamefont {Weniger}}]{Bringmann:2012vr}%
  \BibitemOpen
  \bibfield  {author} {\bibinfo {author} {\bibfnamefont {T.}~\bibnamefont
  {Bringmann}}, \bibinfo {author} {\bibfnamefont {X.}~\bibnamefont {Huang}},
  \bibinfo {author} {\bibfnamefont {A.}~\bibnamefont {Ibarra}}, \bibinfo
  {author} {\bibfnamefont {S.}~\bibnamefont {Vogl}}, \ and\ \bibinfo {author}
  {\bibfnamefont {C.}~\bibnamefont {Weniger}},\ }\href@noop {} {\  (\bibinfo
  {year} {2012})},\ \Eprint {http://arxiv.org/abs/1203.1312} {arXiv:1203.1312
  [hep-ph]} \BibitemShut {NoStop}%
%%CITATION = ARXIV:1203.1312;%%
\bibitem [{\citenamefont {Weniger}(2012)}]{Weniger:2012tx}%
  \BibitemOpen
  \bibfield  {author} {\bibinfo {author} {\bibfnamefont {C.}~\bibnamefont
  {Weniger}},\ }\href@noop {} {\  (\bibinfo {year} {2012})},\ \Eprint
  {http://arxiv.org/abs/1204.2797} {arXiv:1204.2797 [hep-ph]} \BibitemShut
  {NoStop}%
%%CITATION = ARXIV:1204.2797;%%
\bibitem [{\citenamefont {Tempel}\ \emph {et~al.}(2012)\citenamefont {Tempel},
  \citenamefont {Hektor},\ and\ \citenamefont {Raidal}}]{Tempel:2012ey}%
  \BibitemOpen
  \bibfield  {author} {\bibinfo {author} {\bibfnamefont {E.}~\bibnamefont
  {Tempel}}, \bibinfo {author} {\bibfnamefont {A.}~\bibnamefont {Hektor}}, \
  and\ \bibinfo {author} {\bibfnamefont {M.}~\bibnamefont {Raidal}},\
  }\href@noop {} {\  (\bibinfo {year} {2012})},\ \Eprint
  {http://arxiv.org/abs/1205.1045} {arXiv:1205.1045 [hep-ph]} \BibitemShut
  {NoStop}%
%%CITATION = ARXIV:1205.1045;%%
\bibitem [{\citenamefont {Su}\ and\ \citenamefont
  {Finkbeiner}(2012)}]{Su:2012ft}%
  \BibitemOpen
  \bibfield  {author} {\bibinfo {author} {\bibfnamefont {M.}~\bibnamefont
  {Su}}\ and\ \bibinfo {author} {\bibfnamefont {D.~P.}\ \bibnamefont
  {Finkbeiner}},\ }\href@noop {} {\  (\bibinfo {year} {2012})},\ \Eprint
  {http://arxiv.org/abs/1206.1616} {arXiv:1206.1616 [astro-ph.HE]} \BibitemShut
  {NoStop}%
%%CITATION = ARXIV:1206.1616;%%
\bibitem [{\citenamefont {Hektor}\ \emph
  {et~al.}(2012{\natexlab{a}})\citenamefont {Hektor}, \citenamefont {Raidal},\
  and\ \citenamefont {Tempel}}]{Hektor:2012jc}%
  \BibitemOpen
  \bibfield  {author} {\bibinfo {author} {\bibfnamefont {A.}~\bibnamefont
  {Hektor}}, \bibinfo {author} {\bibfnamefont {M.}~\bibnamefont {Raidal}}, \
  and\ \bibinfo {author} {\bibfnamefont {E.}~\bibnamefont {Tempel}},\
  }\href@noop {} {\  (\bibinfo {year} {2012}{\natexlab{a}})},\ \Eprint
  {http://arxiv.org/abs/1208.1996} {arXiv:1208.1996 [astro-ph.HE]} \BibitemShut
  {NoStop}%
%%CITATION = ARXIV:1208.1996;%%
\bibitem [{\citenamefont {Hektor}\ \emph {et~al.}(2013)\citenamefont {Hektor},
  \citenamefont {Raidal},\ and\ \citenamefont {Tempel}}]{Hektor:2012kc}%
  \BibitemOpen
  \bibfield  {author} {\bibinfo {author} {\bibfnamefont {A.}~\bibnamefont
  {Hektor}}, \bibinfo {author} {\bibfnamefont {M.}~\bibnamefont {Raidal}}, \
  and\ \bibinfo {author} {\bibfnamefont {E.}~\bibnamefont {Tempel}},\ }\href
  {\doibase 10.1088/2041-8205/762/2/L22} {\bibfield  {journal} {\bibinfo
  {journal} {Astrophys.J.}\ }\textbf {\bibinfo {volume} {762}},\ \bibinfo
  {pages} {L22} (\bibinfo {year} {2013})},\ \Eprint
  {http://arxiv.org/abs/1207.4466} {arXiv:1207.4466 [astro-ph.HE]} \BibitemShut
  {NoStop}%
%%CITATION = ARXIV:1207.4466;%%
\bibitem [{\citenamefont {Hektor}\ \emph
  {et~al.}(2012{\natexlab{b}})\citenamefont {Hektor}, \citenamefont {Raidal},\
  and\ \citenamefont {Tempel}}]{Hektor:2012ev}%
  \BibitemOpen
  \bibfield  {author} {\bibinfo {author} {\bibfnamefont {A.}~\bibnamefont
  {Hektor}}, \bibinfo {author} {\bibfnamefont {M.}~\bibnamefont {Raidal}}, \
  and\ \bibinfo {author} {\bibfnamefont {E.}~\bibnamefont {Tempel}},\
  }\href@noop {} {\  (\bibinfo {year} {2012}{\natexlab{b}})},\ \Eprint
  {http://arxiv.org/abs/1209.4548} {arXiv:1209.4548 [astro-ph.HE]} \BibitemShut
  {NoStop}%
%%CITATION = ARXIV:1209.4548;%%
\bibitem [{\citenamefont {Whiteson}(2012)}]{Whiteson:2012hr}%
  \BibitemOpen
  \bibfield  {author} {\bibinfo {author} {\bibfnamefont {D.}~\bibnamefont
  {Whiteson}},\ }\href {\doibase 10.1088/1475-7516/2012/11/008} {\bibfield
  {journal} {\bibinfo  {journal} {JCAP}\ }\textbf {\bibinfo {volume} {1211}},\
  \bibinfo {pages} {008} (\bibinfo {year} {2012})},\ \Eprint
  {http://arxiv.org/abs/1208.3677} {arXiv:1208.3677 [astro-ph.HE]} \BibitemShut
  {NoStop}%
%%CITATION = ARXIV:1208.3677;%%
\bibitem [{\citenamefont {Finkbeiner}\ \emph {et~al.}(2013)\citenamefont
  {Finkbeiner}, \citenamefont {Su},\ and\ \citenamefont
  {Weniger}}]{Finkbeiner:2012ez}%
  \BibitemOpen
  \bibfield  {author} {\bibinfo {author} {\bibfnamefont {D.~P.}\ \bibnamefont
  {Finkbeiner}}, \bibinfo {author} {\bibfnamefont {M.}~\bibnamefont {Su}}, \
  and\ \bibinfo {author} {\bibfnamefont {C.}~\bibnamefont {Weniger}},\ }\href
  {\doibase 10.1088/1475-7516/2013/01/029} {\bibfield  {journal} {\bibinfo
  {journal} {JCAP}\ }\textbf {\bibinfo {volume} {1301}},\ \bibinfo {pages}
  {029} (\bibinfo {year} {2013})},\ \Eprint {http://arxiv.org/abs/1209.4562}
  {arXiv:1209.4562 [astro-ph.HE]} \BibitemShut {NoStop}%
%%CITATION = ARXIV:1209.4562;%%
\bibitem [{\citenamefont {Rao}\ and\ \citenamefont
  {Whiteson}(2012)}]{Rao:2012fh}%
  \BibitemOpen
  \bibfield  {author} {\bibinfo {author} {\bibfnamefont {K.}~\bibnamefont
  {Rao}}\ and\ \bibinfo {author} {\bibfnamefont {D.}~\bibnamefont {Whiteson}},\
  }\href@noop {} {\  (\bibinfo {year} {2012})},\ \Eprint
  {http://arxiv.org/abs/1210.4934} {arXiv:1210.4934 [astro-ph.HE]} \BibitemShut
  {NoStop}%
%%CITATION = ARXIV:1210.4934;%%
\bibitem [{\citenamefont {Buchmuller}\ and\ \citenamefont
  {Garny}(2012)}]{Buchmuller:2012rc}%
  \BibitemOpen
  \bibfield  {author} {\bibinfo {author} {\bibfnamefont {W.}~\bibnamefont
  {Buchmuller}}\ and\ \bibinfo {author} {\bibfnamefont {M.}~\bibnamefont
  {Garny}},\ }\href {\doibase 10.1088/1475-7516/2012/08/035} {\bibfield
  {journal} {\bibinfo  {journal} {JCAP}\ }\textbf {\bibinfo {volume} {1208}},\
  \bibinfo {pages} {035} (\bibinfo {year} {2012})},\ \Eprint
  {http://arxiv.org/abs/1206.7056} {arXiv:1206.7056 [hep-ph]} \BibitemShut
  {NoStop}%
%%CITATION = ARXIV:1206.7056;%%
\bibitem [{\citenamefont {Cohen}\ \emph {et~al.}(2012)\citenamefont {Cohen},
  \citenamefont {Lisanti}, \citenamefont {Slatyer},\ and\ \citenamefont
  {Wacker}}]{Cohen:2012me}%
  \BibitemOpen
  \bibfield  {author} {\bibinfo {author} {\bibfnamefont {T.}~\bibnamefont
  {Cohen}}, \bibinfo {author} {\bibfnamefont {M.}~\bibnamefont {Lisanti}},
  \bibinfo {author} {\bibfnamefont {T.~R.}\ \bibnamefont {Slatyer}}, \ and\
  \bibinfo {author} {\bibfnamefont {J.~G.}\ \bibnamefont {Wacker}},\ }\href
  {\doibase 10.1007/JHEP10(2012)134} {\bibfield  {journal} {\bibinfo  {journal}
  {JHEP}\ }\textbf {\bibinfo {volume} {1210}},\ \bibinfo {pages} {134}
  (\bibinfo {year} {2012})},\ \Eprint {http://arxiv.org/abs/1207.0800}
  {arXiv:1207.0800 [hep-ph]} \BibitemShut {NoStop}%
%%CITATION = ARXIV:1207.0800;%%
\bibitem [{\citenamefont {Cholis}\ \emph {et~al.}(2012)\citenamefont {Cholis},
  \citenamefont {Tavakoli},\ and\ \citenamefont {Ullio}}]{Cholis:2012fb}%
  \BibitemOpen
  \bibfield  {author} {\bibinfo {author} {\bibfnamefont {I.}~\bibnamefont
  {Cholis}}, \bibinfo {author} {\bibfnamefont {M.}~\bibnamefont {Tavakoli}}, \
  and\ \bibinfo {author} {\bibfnamefont {P.}~\bibnamefont {Ullio}},\ }\href
  {\doibase 10.1103/PhysRevD.86.083525} {\bibfield  {journal} {\bibinfo
  {journal} {Phys.Rev.}\ }\textbf {\bibinfo {volume} {D86}},\ \bibinfo {pages}
  {083525} (\bibinfo {year} {2012})},\ \Eprint {http://arxiv.org/abs/1207.1468}
  {arXiv:1207.1468 [hep-ph]} \BibitemShut {NoStop}%
%%CITATION = ARXIV:1207.1468;%%
\bibitem [{\citenamefont {Blanchet}\ and\ \citenamefont
  {Lavalle}(2012)}]{Blanchet:2012vq}%
  \BibitemOpen
  \bibfield  {author} {\bibinfo {author} {\bibfnamefont {S.}~\bibnamefont
  {Blanchet}}\ and\ \bibinfo {author} {\bibfnamefont {J.}~\bibnamefont
  {Lavalle}},\ }\href {\doibase 10.1088/1475-7516/2012/11/021} {\bibfield
  {journal} {\bibinfo  {journal} {JCAP}\ }\textbf {\bibinfo {volume} {1211}},\
  \bibinfo {pages} {021} (\bibinfo {year} {2012})},\ \Eprint
  {http://arxiv.org/abs/1207.2476} {arXiv:1207.2476 [astro-ph.HE]} \BibitemShut
  {NoStop}%
%%CITATION = ARXIV:1207.2476;%%
\bibitem [{\citenamefont {Asano}\ \emph {et~al.}(2012)\citenamefont {Asano},
  \citenamefont {Bringmann}, \citenamefont {Sigl},\ and\ \citenamefont
  {Vollmann}}]{Asano:2012zv}%
  \BibitemOpen
  \bibfield  {author} {\bibinfo {author} {\bibfnamefont {M.}~\bibnamefont
  {Asano}}, \bibinfo {author} {\bibfnamefont {T.}~\bibnamefont {Bringmann}},
  \bibinfo {author} {\bibfnamefont {G.}~\bibnamefont {Sigl}}, \ and\ \bibinfo
  {author} {\bibfnamefont {M.}~\bibnamefont {Vollmann}},\ }\href@noop {} {\
  (\bibinfo {year} {2012})},\ \Eprint {http://arxiv.org/abs/1211.6739}
  {arXiv:1211.6739 [hep-ph]} \BibitemShut {NoStop}%
%%CITATION = ARXIV:1211.6739;%%
\bibitem [{\citenamefont {Weiner}\ and\ \citenamefont
  {Yavin}(2012)}]{Weiner:2012cb}%
  \BibitemOpen
  \bibfield  {author} {\bibinfo {author} {\bibfnamefont {N.}~\bibnamefont
  {Weiner}}\ and\ \bibinfo {author} {\bibfnamefont {I.}~\bibnamefont {Yavin}},\
  }\href {\doibase 10.1103/PhysRevD.86.075021} {\bibfield  {journal} {\bibinfo
  {journal} {Phys.Rev.}\ }\textbf {\bibinfo {volume} {D86}},\ \bibinfo {pages}
  {075021} (\bibinfo {year} {2012})},\ \Eprint {http://arxiv.org/abs/1206.2910}
  {arXiv:1206.2910 [hep-ph]} \BibitemShut {NoStop}%
%%CITATION = ARXIV:1206.2910;%%
\bibitem [{\citenamefont {Tulin}\ \emph {et~al.}(2012)\citenamefont {Tulin},
  \citenamefont {Yu},\ and\ \citenamefont {Zurek}}]{Tulin:2012uq}%
  \BibitemOpen
  \bibfield  {author} {\bibinfo {author} {\bibfnamefont {S.}~\bibnamefont
  {Tulin}}, \bibinfo {author} {\bibfnamefont {H.-B.}\ \bibnamefont {Yu}}, \
  and\ \bibinfo {author} {\bibfnamefont {K.~M.}\ \bibnamefont {Zurek}},\
  }\href@noop {} {\  (\bibinfo {year} {2012})},\ \Eprint
  {http://arxiv.org/abs/1208.0009} {arXiv:1208.0009 [hep-ph]} \BibitemShut
  {NoStop}%
%%CITATION = ARXIV:1208.0009;%%
\bibitem [{\citenamefont {Cline}\ \emph {et~al.}(2012)\citenamefont {Cline},
  \citenamefont {Moore},\ and\ \citenamefont {Frey}}]{Cline:2012bz}%
  \BibitemOpen
  \bibfield  {author} {\bibinfo {author} {\bibfnamefont {J.~M.}\ \bibnamefont
  {Cline}}, \bibinfo {author} {\bibfnamefont {G.~D.}\ \bibnamefont {Moore}}, \
  and\ \bibinfo {author} {\bibfnamefont {A.~R.}\ \bibnamefont {Frey}},\ }\href
  {\doibase 10.1103/PhysRevD.86.115013} {\bibfield  {journal} {\bibinfo
  {journal} {Phys.Rev.}\ }\textbf {\bibinfo {volume} {D86}},\ \bibinfo {pages}
  {115013} (\bibinfo {year} {2012})},\ \Eprint {http://arxiv.org/abs/1208.2685}
  {arXiv:1208.2685 [hep-ph]} \BibitemShut {NoStop}%
%%CITATION = ARXIV:1208.2685;%%
\bibitem [{\citenamefont {Dissauer}\ \emph {et~al.}(2013)\citenamefont
  {Dissauer}, \citenamefont {Frandsen}, \citenamefont {Hapola},\ and\
  \citenamefont {Sannino}}]{Dissauer:2012xa}%
  \BibitemOpen
  \bibfield  {author} {\bibinfo {author} {\bibfnamefont {K.}~\bibnamefont
  {Dissauer}}, \bibinfo {author} {\bibfnamefont {M.~T.}\ \bibnamefont
  {Frandsen}}, \bibinfo {author} {\bibfnamefont {T.}~\bibnamefont {Hapola}}, \
  and\ \bibinfo {author} {\bibfnamefont {F.}~\bibnamefont {Sannino}},\ }\href
  {\doibase 10.1103/PhysRevD.87.035005} {\bibfield  {journal} {\bibinfo
  {journal} {Phys.Rev.}\ }\textbf {\bibinfo {volume} {D87}},\ \bibinfo {pages}
  {035005} (\bibinfo {year} {2013})},\ \Eprint {http://arxiv.org/abs/1211.5144}
  {arXiv:1211.5144 [hep-ph]} \BibitemShut {NoStop}%
%%CITATION = ARXIV:1211.5144;%%
\bibitem [{\citenamefont {Weiner}\ and\ \citenamefont
  {Yavin}(2013)}]{Weiner:2012gm}%
  \BibitemOpen
  \bibfield  {author} {\bibinfo {author} {\bibfnamefont {N.}~\bibnamefont
  {Weiner}}\ and\ \bibinfo {author} {\bibfnamefont {I.}~\bibnamefont {Yavin}},\
  }\href {\doibase 10.1103/PhysRevD.87.023523} {\bibfield  {journal} {\bibinfo
  {journal} {Phys.Rev.}\ }\textbf {\bibinfo {volume} {D87}},\ \bibinfo {pages}
  {023523} (\bibinfo {year} {2013})},\ \Eprint {http://arxiv.org/abs/1209.1093}
  {arXiv:1209.1093 [hep-ph]} \BibitemShut {NoStop}%
%%CITATION = ARXIV:1209.1093;%%
\bibitem [{\citenamefont {Liu}\ \emph {et~al.}(2013)\citenamefont {Liu},
  \citenamefont {Shuve}, \citenamefont {Weiner},\ and\ \citenamefont
  {Yavin}}]{Liu:2013gba}%
  \BibitemOpen
  \bibfield  {author} {\bibinfo {author} {\bibfnamefont {J.}~\bibnamefont
  {Liu}}, \bibinfo {author} {\bibfnamefont {B.}~\bibnamefont {Shuve}}, \bibinfo
  {author} {\bibfnamefont {N.}~\bibnamefont {Weiner}}, \ and\ \bibinfo {author}
  {\bibfnamefont {I.}~\bibnamefont {Yavin}},\ }\href@noop {} {\  (\bibinfo
  {year} {2013})},\ \Eprint {http://arxiv.org/abs/1303.4404} {arXiv:1303.4404
  [hep-ph]} \BibitemShut {NoStop}%
%%CITATION = ARXIV:1303.4404;%%
\bibitem [{\citenamefont {Borowka}\ \emph {et~al.}(2013)\citenamefont
  {Borowka}, \citenamefont {Carter},\ and\ \citenamefont
  {Heinrich}}]{Borowka:2012yc}%
  \BibitemOpen
  \bibfield  {author} {\bibinfo {author} {\bibfnamefont {S.}~\bibnamefont
  {Borowka}}, \bibinfo {author} {\bibfnamefont {J.}~\bibnamefont {Carter}}, \
  and\ \bibinfo {author} {\bibfnamefont {G.}~\bibnamefont {Heinrich}},\ }\href
  {\doibase 10.1016/j.cpc.2012.09.020} {\bibfield  {journal} {\bibinfo
  {journal} {Comput.Phys.Commun.}\ }\textbf {\bibinfo {volume} {184}},\
  \bibinfo {pages} {396} (\bibinfo {year} {2013})},\ \Eprint
  {http://arxiv.org/abs/1204.4152} {arXiv:1204.4152 [hep-ph]} \BibitemShut
  {NoStop}%
%%CITATION = ARXIV:1204.4152;%%
\bibitem [{\citenamefont {Barger}\ \emph {et~al.}(2010)\citenamefont {Barger},
  \citenamefont {Keung},\ and\ \citenamefont {Marfatia}}]{Barger:2010gv}%
  \BibitemOpen
  \bibfield  {author} {\bibinfo {author} {\bibfnamefont {V.}~\bibnamefont
  {Barger}}, \bibinfo {author} {\bibfnamefont {W.-Y.}\ \bibnamefont {Keung}}, \
  and\ \bibinfo {author} {\bibfnamefont {D.}~\bibnamefont {Marfatia}},\
  }\href@noop {} {\  (\bibinfo {year} {2010})},\ \Eprint
  {http://arxiv.org/abs/1007.4345} {arXiv:1007.4345 [hep-ph]} \BibitemShut
  {NoStop}%
%%CITATION = 1007.4345;%%
\bibitem [{\citenamefont {Bernabei}\ \emph {et~al.}(2010)\citenamefont
  {Bernabei} \emph {et~al.}}]{Bernabei:2010mq}%
  \BibitemOpen
  \bibfield  {author} {\bibinfo {author} {\bibfnamefont {R.}~\bibnamefont
  {Bernabei}} \emph {et~al.},\ }\href {\doibase 10.1140/epjc/s10052-010-1303-9}
  {\  (\bibinfo {year} {2010}),\ 10.1140/epjc/s10052-010-1303-9},\ \Eprint
  {http://arxiv.org/abs/1002.1028} {arXiv:1002.1028 [astro-ph.GA]} \BibitemShut
  {NoStop}%
%%CITATION = 1002.1028;%%
\bibitem [{\citenamefont {Pradler}\ \emph {et~al.}(2013)\citenamefont
  {Pradler}, \citenamefont {Singh},\ and\ \citenamefont
  {Yavin}}]{Pradler:2012qt}%
  \BibitemOpen
  \bibfield  {author} {\bibinfo {author} {\bibfnamefont {J.}~\bibnamefont
  {Pradler}}, \bibinfo {author} {\bibfnamefont {B.}~\bibnamefont {Singh}}, \
  and\ \bibinfo {author} {\bibfnamefont {I.}~\bibnamefont {Yavin}},\ }\href
  {\doibase 10.1016/j.physletb.2013.02.033} {\bibfield  {journal} {\bibinfo
  {journal} {Phys.Lett.}\ }\textbf {\bibinfo {volume} {B720}},\ \bibinfo
  {pages} {399} (\bibinfo {year} {2013})},\ \Eprint
  {http://arxiv.org/abs/1210.5501} {arXiv:1210.5501 [hep-ph]} \BibitemShut
  {NoStop}%
%%CITATION = ARXIV:1210.5501;%%
\bibitem [{\citenamefont {Pradler}\ and\ \citenamefont
  {Yavin}(2012)}]{Pradler:2012bf}%
  \BibitemOpen
  \bibfield  {author} {\bibinfo {author} {\bibfnamefont {J.}~\bibnamefont
  {Pradler}}\ and\ \bibinfo {author} {\bibfnamefont {I.}~\bibnamefont
  {Yavin}},\ }\href@noop {} {\  (\bibinfo {year} {2012})},\ \Eprint
  {http://arxiv.org/abs/1210.7548} {arXiv:1210.7548 [hep-ph]} \BibitemShut
  {NoStop}%
%%CITATION = ARXIV:1210.7548;%%
\bibitem [{\citenamefont {Martin}\ \emph {et~al.}(2009)\citenamefont {Martin},
  \citenamefont {Stirling}, \citenamefont {Thorne},\ and\ \citenamefont
  {Watt}}]{Martin:2009iq}%
  \BibitemOpen
  \bibfield  {author} {\bibinfo {author} {\bibfnamefont {A.}~\bibnamefont
  {Martin}}, \bibinfo {author} {\bibfnamefont {W.}~\bibnamefont {Stirling}},
  \bibinfo {author} {\bibfnamefont {R.}~\bibnamefont {Thorne}}, \ and\ \bibinfo
  {author} {\bibfnamefont {G.}~\bibnamefont {Watt}},\ }\href {\doibase
  10.1140/epjc/s10052-009-1072-5} {\bibfield  {journal} {\bibinfo  {journal}
  {Eur.Phys.J.}\ }\textbf {\bibinfo {volume} {C63}},\ \bibinfo {pages} {189}
  (\bibinfo {year} {2009})},\ \Eprint {http://arxiv.org/abs/0901.0002}
  {arXiv:0901.0002 [hep-ph]} \BibitemShut {NoStop}%
%%CITATION = ARXIV:0901.0002;%%
\bibitem [{\citenamefont {Chatrchyan}\ \emph {et~al.}(2012)\citenamefont
  {Chatrchyan} \emph {et~al.}}]{Chatrchyan:2012me}%
  \BibitemOpen
  \bibfield  {author} {\bibinfo {author} {\bibfnamefont {S.}~\bibnamefont
  {Chatrchyan}} \emph {et~al.} (\bibinfo {collaboration} {CMS Collaboration}),\
  }\href {\doibase 10.1007/JHEP09(2012)094} {\bibfield  {journal} {\bibinfo
  {journal} {JHEP}\ }\textbf {\bibinfo {volume} {1209}},\ \bibinfo {pages}
  {094} (\bibinfo {year} {2012})},\ \Eprint {http://arxiv.org/abs/1206.5663}
  {arXiv:1206.5663 [hep-ex]} \BibitemShut {NoStop}%
%%CITATION = ARXIV:1206.5663;%%
\bibitem [{CMS()}]{CMS:rwa}%
  \BibitemOpen
  \href@noop {} {\ }\Eprint {http://arxiv.org/abs/CMS Collaboration,
  CMS-PAS-EXO-12-048} {CMS Collaboration, CMS-PAS-EXO-12-048} \BibitemShut
  {NoStop}%
\end{thebibliography}%

%%%%%%%%%%%%%%%%
% Appendix B
%%%%%%%%%%%%%%%%

\end{document}